\documentclass[preprint,12pt]{article}

\usepackage{amsmath,amssymb,array,calc,rotating,epsfig,psfrag, amscd, datetime}
\numberwithin{equation}{section}

\usepackage{color}
\usepackage[
      colorlinks=true,
      urlcolor=blue,    
      filecolor=blue,     
      citecolor=red,
      pdfstartview=FitV,
       bookmarksopen=true    
      ]{hyperref}
\usepackage[left=2cm,top=2.5cm,right=2cm,nohead]{geometry}

\newcommand{\nc}{\newcommand}

\definecolor{cardinal}{rgb}{0.6,0,0}
\definecolor{darkgreen}{rgb}{0,0.5,0}
\definecolor{golden}{rgb}{0.92, 0.7, 0}
\definecolor{midnight}{rgb}{0, 0, 0.5}
\definecolor{darkblue}{rgb}{0.2, 0, 0.8}

\nc{\ra}{\rightarrow} 
\nc{\lra}{\leftrightarrow} 
\nc{\Ra}{\Rightarrow} 
\nc{\LRa}{\Leftightarrow} 
\nc{\blp}{{\big (}}
\nc{\brp}{{\big )}}
\nc{\Blp}{{\Big (}}
\nc{\Brp}{{\Big )}}
\nc{\bglp}{{\bigg (}}
\nc{\bgrp}{{\bigg )}}
\nc{\Bglp}{{\Bigg (}}
\nc{\Bgrp}{{\Bigg )}}
\nc{\slb}{{\rm [}}
\nc{\srb}{{\rm ]}}
\nc{\bslb}{{\rm \big [}}
\nc{\bsrb}{{\rm \big ]}}
\nc{\Bslb}{{\rm \Big [}}
\nc{\Bsrb}{{\rm \Big ]}}

\def\al{\alpha}

\def\eps{\epsilon}
\nc{\veps}{\varepsilon}
\def\gam{\gamma}

\def\lam{\lambda}
\def\om{\omega}

\nc{\vphi}{\varphi}
\def\tha{\theta}

\def\sig{\sigma}

\def\Gam{\Gamma}

\def\Lam{\Lambda}
\def\Om{\Omega}
\def\Sig{\Sigma}

\def\coeff#1#2{\relax{\textstyle {#1 \over #2}}\displaystyle}

\nc{\myvspace}{\rule[-1em]{0pt}{2.5em}}
\nc{\bea}{\begin{eqnarray}}
\nc{\eea}{\end{eqnarray}}
\nc{\be}{\begin{equation}}
\nc{\ee}{\end{equation}}
\nc{\barr}{\begin{array}}
\nc{\earr}{\end{array}}

\nc{\co}{{\cal o}}

\nc{\cA}{{\cal A}}
\nc{\cB}{{ \cal B}}

\def\cD{{\cal D}}

\nc{\cF}{{\cal F}}
\nc{\cG}{{\cal G}}

\def\cI{{\cal I}}

\def\cK{{\cal K}}
\nc{\cL}{{\cal L}}
\nc{\cM}{{\cal M}}
\def\N{{\cal N}}
\def\cN{{\cal N}}
\def\cNbar{\ol{\cal N}}
\def\cO{{\cal O}}

\nc{\cQ}{{\cal Q}}
\nc{\cR}{{\cal R}}
\def\cS{{\cal S}}

\def\cV{{\cal V}}
\def\cV{{\cal V}}
\def\cW{{\cal W}}

\def\cZ{{\cal Z}}
\nc{\cQd}{\cQ^{\dagger}}
\nc{\cRd}{\cR^{\dagger}}
\nc{\BB}{{\mathbb B}}
\nc{\CC}{{\mathbb C}}
\nc{\DD}{{\mathbb D}}
\nc{\EE}{{\mathbb E}}
\nc{\FF}{{\mathbb F}}
\nc{\GG}{{\mathbb G}}
\nc{\HH}{{\mathbb H}}
\nc{\JJ}{{\mathbb J}}
\nc{\MM}{{\mathbb M}}
\nc{\RR}{{\mathbb R}}
\nc{\PP}{{\mathbb P}}
\nc{\QQ}{{\mathbb Q}}
\nc{\UU}{{\mathbb U}}
\nc{\ZZ}{{\mathbb Z}}
\nc{\calone}{{\mathbb 1}}

\nc{\half}{\coeff{1}{2}}
\nc{\quarter}{\coeff{1}{4}}
\nc{\del}{\partial}

\nc{\delbar}{\bar\partial}
\nc{\thalf}{\frac{t}{2}}
\nc{\Spin}{\operatorname{Spin}}
\nc{\SO}{\operatorname{SO}}

\nc{\Sp}{{\rm Sp}}
\nc{\com}[2]{{ \left[ #1, #2 \right] }}
\nc{\acom}[2]{{ \left\{ #1, #2 \right\} }}
\nc{\rr}{\rightarrow}
\nc{\p}{\partial}
\nc{\LT}{{\LL_\T}}
\nc{\Tr}{{\rm Tr}}
\nc{\tr}{{\rm tr}}
\nc{\Adag}{A^{\dagger}}
\nc{\AdagI}{A^{\dagger I}}
\nc{\AdagJ}{A^{\dagger J}}
\nc{\AdagK}{A^{\dagger K}}
\nc{\AdagL}{A^{\dagger L}}
\nc{\AdagM}{A^{\dagger M}}
\nc{\Bdag}{B^{\dagger}}
\nc{\BdagI}{B^{\dagger}_I}
\nc{\BdagJ}{B^{\dagger}_J}
\nc{\BdagK}{B^{\dagger}_K}
\nc{\BdagL}{B^{\dagger}_L}
\nc{\BdagM}{B^{\dagger}_M}
\nc{\Cdag}{C^{\dagger}}
\nc{\CdagI}{C^{\dagger I}}
\nc{\CdagJ}{C^{\dagger J}}
\nc{\CdagK}{C^{\dagger K}}
\nc{\Ddag}{D^{\dagger}}
\nc{\DdagI}{D^{\dagger I}}
\nc{\DdagJ}{D^{\dagger J}}
\nc{\DdagK}{D^{\dagger K}}
\nc{\bva}{\breve{a}}
\nc{\bvb}{\breve{b}}
\nc{\bvc}{\breve{c}}
\nc{\bvd}{\breve{d}}
\nc{\bve}{\breve{e}}
\nc{\bvf}{\breve{f}}
\nc{\bvg}{\breve{g}}
\nc{\bvh}{\breve{h}}
\nc{\bvi}{\breve{i}}
\nc{\bvj}{\breve{j}}
\nc{\bvk}{\breve{k}}
\nc{\bvl}{\breve{l}}
\nc{\bvm}{\breve{m}}
\nc{\bvn}{\breve{n}}
\nc{\bvo}{\breve{o}}
\nc{\bvp}{\breve{p}}
\nc{\brvq}{\breve{q}}
\nc{\bvr}{\breve{r}}
\nc{\bvs}{\breve{s}}
\nc{\bvt}{\breve{t}}
\nc{\bvu}{\breve{u}}
\nc{\bvv}{\breve{v}}
\nc{\bvw}{\breve{w}}
\nc{\bvx}{\breve{x}}
\nc{\bvy}{\breve{y}}
\nc{\bvz}{\breve{z}}

\nc{\bvA}{\breve{A}}
\nc{\bvB}{\breve{B}}
\nc{\bvC}{\breve{C}}
\nc{\bvD}{\breve{D}}
\nc{\bvE}{\breve{E}}
\nc{\bvF}{\breve{F}}
\nc{\bvG}{\breve{G}}
\nc{\bvH}{\breve{H}}
\nc{\bvI}{\breve{I}}
\nc{\bvJ}{\breve{J}}
\nc{\bvK}{\breve{K}}
\nc{\bvL}{\breve{L}}
\nc{\bvM}{\breve{M}}
\nc{\bvN}{\breve{N}}
\nc{\bvO}{\breve{O}}
\nc{\bvP}{\breve{P}}
\nc{\bvQ}{\breve{Q}}
\nc{\bvR}{\breve{R}}
\nc{\bvS}{\breve{S}}
\nc{\bvT}{\breve{T}}
\nc{\bvU}{\breve{U}}
\nc{\bvV}{\breve{V}}
\nc{\bvcV}{\breve{\cV}}
\nc{\bvW}{\breve{W}}
\nc{\bvX}{\breve{X}}
\nc{\bvY}{\breve{Y}}
\nc{\bvZ}{\breve{Z}}

\nc{\ul}[1]{{\underline{#1}}}

\nc{\tal}{\widetilde{\alpha}}
\nc{\tbeta}{\widetilde{\beta}}
\nc{\ttha}{\tilde{\theta}}
\nc{\ttau}{\tilde{\tau}}
\nc{\tTha}{\tilde{\Theta}}
\nc{\tphi}{\tilde{\phi}}
\nc{\tsig}{\tilde{\sig}}
\nc{\tom}{\widetilde{\om}}
\nc{\tOm}{\widetilde{\Om}}
\nc{\tlam}{\widetilde{\lam}}
\nc{\tLam}{\tilde{\Lam}}
\nc{\tSig}{\widetilde{\Sig}}
\nc{\tPhi}{\tilde{\Phi}}
\nc{\tPhibar}{\ol{\tPhi}}
\nc{\tPi}{\widetilde{\Pi}}
\nc{\tpsi}{\widetilde{\psi}}
\nc{\tPsi}{\tilde{\Psi}}
\nc{\tgam}{\widetilde{\gam}}
\nc{\tGam}{\widetilde{\Gam}}
\nc{\tzeta}{\tilde{\zeta}}
\nc{\tZeta}{\tilde{\Zeta}}
\nc{\teta}{\widetilde{\eta}}
\nc{\teps}{\tilde{\eps}}
\nc{\tveps}{\tilde{\veps}}
\nc{\tEta}{\tilde{\Eta}}
\nc{\tchi}{\tilde{\chi}}
\nc{\tChi}{\tilde{\Chi}}
\nc{\txi}{\tilde{\xi}}
\nc{\tXi}{\widetilde{\Xi}}
\nc{\tnu}{\tilde{\nu}}
\nc{\tmu}{\tilde{\mu}}

\nc{\ta}{\tilde a}
\nc{\tb}{\tilde b}
\nc{\tc}{\tilde c}
\nc{\te}{\tilde e}
\nc{\tf}{\widetilde f}
\nc{\tg}{\widetilde g}
\nc{\ti}{\tilde i}
\nc{\tj}{\tilde j}
\nc{\tk}{\widetilde k}
\nc{\tl}{\tilde l}
\nc{\tm}{\widetilde m}
\nc{\tn}{\tilde n}
\nc{\tp}{\tilde{p}}
\nc{\tq}{\widetilde{q}}
\nc{\trr}{{\tilde r}}
\nc{\ts}{{\tilde s}}
\nc{\tu}{{\tilde u}}
\nc{\tv}{{\tilde v}}
\nc{\tw}{{\tilde w}}
\nc{\tx}{{\tilde x}}
\nc{\ty}{{\tilde y}}
\nc{\tz}{\tilde z}
\nc{\tA}{{\widetilde A}}
\nc{\tAbar}{{\ol \tA}}
\nc{\tB}{{\widetilde B}}
\nc{\tC}{{\widetilde C}}
\nc{\tD}{{\widetilde D}}
\nc{\tE}{{\widetilde E}}
\nc{\tF}{{\widetilde F}}
\nc{\tG}{{\widetilde G}}
\nc{\tcG}{{\widetilde \cG}}
\nc{\tH}{{\widetilde H}}
\nc{\tI}{{\widetilde I}}
\nc{\tcI}{{\widetilde \cI}}
\nc{\tJ}{{\widetilde J}}
\nc{\tJbar}{{\ol {\tilde J}}}
\nc{\tK}{{\widetilde K}}
\nc{\tL}{{\widetilde L}}
\nc{\tcL}{{\widetilde \cL}}
\nc{\tcLbar}{{\ol \tcL}}
\nc{\tM}{{\widetilde M}}
\nc{\tN}{{\widetilde N}}
\nc{\tcN}{{\widetilde \cN}}
\nc{\tP}{{\widetilde P}}
\nc{\tQ}{{\widetilde Q}}
\nc{\tR}{{\widetilde R}}
\nc{\tS}{\widetilde{S}}
\nc{\tT}{\widetilde{T}}
\nc{\tU}{\widetilde{U}}
\nc{\tUU}{\widetilde{\UU}}
\nc{\tV}{\widetilde{V}}
\nc{\tcV}{\widetilde{\cV}}
\nc{\tcVbar}{\ol{\widetilde{\cV}}}
\nc{\tW}{\widetilde{W}}
\nc{\tcF}{\widetilde{{\cal F}}}
\nc{\tX}{\widetilde{X}}
\nc{\tY}{\widetilde{Y}}
\nc{\tcZ}{\tilde{\cZ}}
\nc{\tcZbar}{\ol{\tcZ}}

\nc{\ha}{\hat a}
\nc{\hb}{\hat b}
\nc{\hc}{\widehat c}
\nc{\hd}{\widehat d}
\nc{\he}{\widehat e}
\nc{\hf}{\widehat f}
\nc{\hg}{\widehat g}
\nc{\hh}{\widehat h}
\nc{\hm}{\widehat m}
\nc{\hn}{\widehat n}
\nc{\hp}{\widehat p}
\nc{\hq}{\widehat q}
\nc{\hr}{\widehat r}
\nc{\hs}{\widehat s}
\nc{\hv}{\widehat v}
\nc{\hw}{\widehat w}
\nc{\hx}{\widehat x}
\nc{\hy}{\widehat y}
\nc{\hz}{\widehat z}
\nc{\zhat}{\hat z}
\nc{\hA}{\widehat{A}}
\nc{\hB}{\widehat{B}}
\nc{\hC}{\widehat{C}}
\nc{\hD}{\widehat{D}}
\nc{\hE}{\widehat{E}}
\nc{\hF}{\widehat{F}}
\nc{\hcF}{\widehat{\cF}}
\nc{\hG}{\widehat{G}}
\nc{\hcG}{\widehat{\cG}}
\nc{\hH}{\widehat{H}}
\nc{\hI}{\widehat{I}}
\nc{\hcI}{\widehat{\cI}}
\nc{\hJ}{\widehat{J}}
\nc{\hK}{\widehat{K}}
\nc{\hL}{\widehat{L}}
\nc{\hcL}{\widehat{\cL}}
\nc{\hM}{\widehat M}
\nc{\hcM}{\widehat{\cM}}
\nc{\hN}{\widehat{N}}
\nc{\hO}{\widehat{O}}
\nc{\hcO}{\widehat{\cO}}
\nc{\hP}{\widehat{P}}
\nc{\hQ}{\widehat{Q}}
\nc{\hcQ}{\widehat{\cQ}}
\nc{\hcR}{\widehat{\cR}}
\nc{\hR}{\widehat{R}}
\nc{\hS}{\widehat{S}}
\nc{\hcS}{\widehat{\cS}}
\nc{\hT}{\widehat{T}}
\nc{\hU}{\widehat{U}}
\nc{\hV}{\widehat V}
\nc{\hcV}{\widehat \cV}
\nc{\hX}{\widehat X}
\nc{\hcZ}{\widehat \cZ}
\nc{\hcZbar}{\ol{\widehat \cZ}}

\nc{\heta}{\widehat{\eta}}
\nc{\hal}{\widehat \alpha}
\nc{\hbeta}{\widehat \beta}
\nc{\hphi}{\widehat{\phi}}
\nc{\hkap}{\hat{\kappa}}
\nc{\hchi}{\widehat{\chi}}
\nc{\hpsi}{\widehat{\psi}}
\nc{\hgam}{\widehat{\gam}}
\nc{\hPhi}{\hat{\Phi}}
\nc{\hPsi}{\hat{\Psi}}
\nc{\hGam}{\hat{\Gam}}
\nc{\omhat}{\widehat{\om}}
\nc{\htha}{\hat{\tha}}
\nc{\hrho}{\widehat{\rho}}
\nc{\hdel}{\widehat{\del}}
\nc{\hnabla}{\widehat{\nabla}}

\nc{\w}{\wedge}

\nc{\vb}{\vec b}
\nc{\vc}{\vec c}
\nc{\vd}{\vec d}
\nc{\ve}{\vec e}
\nc{\vf}{\vec f}
\nc{\vg}{\vec g}
\nc{\vh}{\vec h}
\nc{\vp}{\vec p}
\nc{\vq}{\vec q}
\nc{\vr}{\vec r}
\nc{\vs}{\vec s}
\nc{\vv}{\vec v}
\nc{\vw}{\vec w}
\nc{\vx}{\vec x}
\nc{\vy}{\vec y}
\nc{\vz}{\vec z}

\nc{\vB}{\vec B}
\nc{\vC}{\vec C}
\nc{\vD}{\vec D}
\nc{\vE}{\vec E}
\nc{\vF}{\vec F}
\nc{\vG}{\vec G}
\nc{\vH}{\vec H}
\nc{\vP}{\vec P}
\nc{\vQ}{\vec Q}
\nc{\vR}{\vec R}
\nc{\vS}{\vec S}
\nc{\vV}{\vec V}
\nc{\vW}{\vec W}
\nc{\vX}{\vec X}
\nc{\vY}{\vec Y}
\nc{\vZ}{\vec Z}

\nc{\ol}{\overline}
\nc{\abar}{\ol{a}}
\nc{\bbar}{\ol{b}}
\nc{\cbar}{\ol{c}}
\nc{\dbar}{\ol{d}}
\nc{\ebar}{\ol{e}}
\nc{\fbar}{\ol{f}}
\nc{\gbar}{\ol{g}}
\nc{\ibar}{\ol{\imath}}
\nc{\jbar}{\ol{\jmath}}
\nc{\kbar}{\ol{k}}
\nc{\lbar}{\ol{l}}
\nc{\mbar}{\ol{m}}
\nc{\nbar}{\ol{n}}
\nc{\pbar}{\ol{p}}
\nc{\qbar}{\ol{q}}
\nc{\rbar}{\ol{r}}
\nc{\sbar}{\ol{s}}
\nc{\ubar}{\ol{u}}
\nc{\vbar}{\ol{v}}
\nc{\wbar}{\ol{w}}
\nc{\xbar}{\ol{x}}
\nc{\ybar}{\ol{y}}
\nc{\zbar}{\ol{z}}

\nc{\Abar}{\ol{A}}
\nc{\Bbar}{\ol{B}}
\nc{\cBbar}{\ol{\cB}}
\nc{\Cbar}{\ol{C}}
\nc{\Dbar}{\ol{D}}
\nc{\Ebar}{\ol{E}}
\nc{\Fbar}{\ol{F}}
\nc{\Gbar}{\ol{G}}
\nc{\Jbar}{\ol{J}}
\nc{\Kbar}{\ol{K}}
\nc{\cKbar}{\ol{\cK}}
\nc{\Lbar}{\ol{L}}
\nc{\cLbar}{\ol{\cL}}
\nc{\Mbar}{\ol{M}}
\nc{\Nbar}{\ol{N}}
\nc{\Pbar}{\ol{P}}
\nc{\Qbar}{\ol{Q}}
\nc{\Rbar}{\ol{R}}
\nc{\Sbar}{\ol{S}}
\nc{\Tbar}{\ol{T}}
\nc{\Ubar}{\ol{U}}
\nc{\Vbar}{\ol{V}}
\nc{\cVbar}{\ol{\cV}}
\nc{\Wbar}{\ol{W}}
\nc{\cWbar}{\ol{\cW}}
\nc{\Xbar}{{\overline X}}
\nc{\Ybar}{{\overline Y}}
\nc{\Zbar}{{\overline Z}}
\nc{\cZbar}{{\overline \cZ}}

\nc{\epsbar}{\ol{\epsilon}}
\nc{\albar}{\ol{\al}}
\nc{\Albar}{\ol{\Al}}
\nc{\betabar}{\ol{\beta}}
\nc{\Betabar}{\ol{\Beta}}
\nc{\lambar}{\ol{\lambda}}
\nc{\kapbar}{\ol{\kappa}}
\nc{\zetabar}{\ol{\zeta}}
\nc{\Zetabar}{\ol{\Zeta}}
\nc{\taubar}{\ol{\tau}}
\nc{\Taubar}{\ol{\Tau}}
\nc{\psibar}{\ol{\psi}}
\nc{\Psibar}{\ol{\Psi}}
\nc{\tpsibar}{\ol{\tpsi}}
\nc{\tPsibar}{\ol{\tPsi}}
\nc{\phibar}{\ol{\phi}}
\nc{\Phibar}{\ol{\Phi}}
\nc{\chibar}{\ol{\chi}}
\nc{\mubar}{\ol{\mu}}
\nc{\nubar}{\ol{\nu}}
\nc{\rhobar}{\ol{\rho}}
\nc{\ombar}{\ol{\om}}
\nc{\Ombar}{\ol{\Om}}
\nc{\Deltabar}{\ol{\Delta}}
\nc{\Thetabar}{\ol{\Theta}}
\nc{\xibar}{\ol{\xi}}
\nc{\Xibar}{\ol{\Xi}}

\nc{\Dthbar}{\ol{\rm D3}}

\nc{\fdot}{\dot{f}}
\nc{\gdot}{\dot{g}}
\nc{\pdot}{\dot{p}}
\nc{\qdot}{\dot{q}}
\nc{\rdot}{\dot{r}}
\nc{\sdot}{\dot{s}}
\nc{\tdot}{\dot{t}}
\nc{\udot}{\dot{u}}
\nc{\vdot}{\dot{v}}
\nc{\wdot}{\dot{w}}
\nc{\xdot}{\dot{x}}
\nc{\xddot}{\ddot{x}}
\nc{\ydot}{\dot{y}}
\nc{\zdot}{\dot{z}}
\nc{\yddot}{\ddot{y}}

\nc{\Adot}{\dot{A}}
\nc{\Bdot}{\dot{B}}
\nc{\Cdot}{\dot{C}}
\nc{\Udot}{\dot{U}}
\nc{\Vdot}{\dot{V}}
\nc{\Wdot}{\dot{W}}

\nc{\taudot}{\dot{\tau}}
\nc{\phidot}{\dot{\phi}}
\nc{\psidot}{\dot{\psi}}
\nc{\chidot}{\dot{\chi}}
\nc{\sinp}{s_{\phi}}
\nc{\cosp}{c_{\phi}}
\nc{\tanp}{t_{\phi}}
\nc{\spone}{s_{\phi_1}}
\nc{\cpone}{c_{\phi_1}}
\nc{\tpone}{t_{\phi_1}}
\nc{\sptwo}{s_{\phi_2}}
\nc{\cptwo}{c_{\phi_2}}
\nc{\tptwo}{t_{\phi_2}}
\nc{\spth}{s_{\phi_3}}
\nc{\cpth}{c_{\phi_3}}
\nc{\tpth}{t_{\phi_3}}
\nc{\calp}{c_{\al}}
\nc{\salp}{s_{\al}}

\nc{\csch}{{\rm csch}}
\nc{\sech}{{\rm sech}}

\nc{\cothzlami}{\coth(z-\lam_i)}
\nc{\coshzlami}{\cosh(z-\lam_i)}
\nc{\sinhzlami}{\sinh(z-\lam_i)}

\nc{\cothzlamj}{\coth(z-\lam_j)}
\nc{\coshzlamj}{\cosh(z-\lam_j)}
\nc{\sinhzlamj}{\sinh(z-\lam_j)}

\nc{\cothlamij}{\coth(\lam_i-\lam_j)}
\nc{\coshlamij}{\cosh(\lam_i-\lam_j)}
\nc{\sinhlamij}{\sinh(\lam_i-\lam_j)}

\nc{\bah}{{\mathbf {\hat{A}}}}
\nc{\bX}{{\mathbf X}}
\nc{\ba}{{\bf a}}
\nc{\bb}{{\bf b}}
\nc{\bc}{{\bf c}}
\nc{\bd}{{\bf d}}
\nc{\bg}{{\bf g}}
\nc{\bk}{{\bf k}}
\nc{\bl}{{\bf l}}
\nc{\bm}{{\bf m}}
\nc{\bn}{{\bf n}}
\nc{\bo}{{\bf o}}
\nc{\bp}{{\bf p}}
\nc{\bq}{{\bf q}}
\nc{\br}{{\bf r}}
\nc{\bs}{{\bf s}}
\nc{\bt}{{\bf t}}
\nc{\bu}{{\bf u}}
\nc{\bv}{{\bf v}}
\nc{\bw}{{\bf w}}
\nc{\bx}{{\bf x}}
\nc{\by}{{\bf y}}
\nc{\bz}{{\bf z}}

\nc{\bP}{{\bf P}}
\nc{\bQ}{{\bf Q}}

\nc{\bom}{{\bf \om}}
\nc{\bombar}{{\mathbf \ombar}}
\nc{\bPhi}{{\bf \Phi}}

\nc{\rma}{{\rm a}}
\nc{\rmb}{{\rm b}}
\nc{\rmc}{{\rm c}}
\nc{\rmd}{{\rm d}}
\nc{\rmg}{{\rm g}}
\nc{\rk}{{\rm k}}
\nc{\rml}{{\rm l}}
\nc{\rmm}{{\rm m}}
\nc{\rmn}{{\rm n}}
\nc{\rmo}{{\rm o}}
\nc{\rmp}{{\rm p}}
\nc{\rmq}{{\rm q}}
\nc{\rmr}{{\rm r}}
\nc{\rms}{{\rm s}}
\nc{\rmt}{{\rm t}}
\nc{\rmu}{{\rm u}}
\nc{\rmv}{{\rm v}}
\nc{\rmw}{{\rm w}}
\nc{\rmx}{{\rm x}}
\nc{\rmy}{{\rm y}}
\nc{\rmz}{{\rm z}}

\nc{\dal}{\dot{\al}}
\nc{\thadot}{\dot{\tha}}
\nc{\thab}{\bar{\theta}}
\nc{\thal}{\theta^{\al}}
\nc{\thdal}{\bar{\theta}^{\dal}}

\nc{\thsigthm}{\tha \sigma^m \thab}
\nc{\thsigthn}{\tha \sigma^n \thab}

\nc{\Dal}{D_{\al}}
\nc{\Ddal}{\bar{D}_{\dal}}
\nc{\CDal}{{\cal D}_{\al}}
\nc{\CDdal}{\bar{\cal D}_{\dal}}

\nc{\eq}[1]{{(\ref{#1})}}
\nc{\eqtwo}[2]{{(\ref{#1},\ref{#2})}}
\nc{\eqthree}[3]{(\ref{#1},\ref{#2},\ref{#3})}
\nc{\eqfour}[4]{(\ref{#1},\ref{#2},\ref{#3},\ref{#4})}
\nc{\eqfive}[5]{(\ref{#1},\ref{#2},\ref{#3},\ref{#4,\ref{#5}})}
\nc{\non}{\nonumber}
\nc{\Fzero}{F_{(0)}}
\nc{\Ftwo}{F_{(2)}}
\nc{\Ffour}{F_{(4)}}
\nc{\Fone}{F_{(1)}}
\nc{\Fthree}{F_{(3)}}
\nc{\Ffive}{F_{(5)}}
\nc{\Fn}{F_{(n)}}
\nc{\Fp}{F_{(p)}}

\nc{\tFzero}{\tF_{(0)}}
\nc{\tFtwo}{\tF_{(2)}}
\nc{\tFfour}{\tF_{(4)}}
\nc{\tFone}{\tF_{(1)}}
\nc{\tFthree}{\tF_{(3)}}
\nc{\tFfive}{\tF_{(5)}}
\nc{\tFn}{\tF_{(n)}}
\nc{\tFp}{\tF_{(p)}}

\nc{\Czero}{C_{(0)}}
\nc{\Ctwo}{C_{(2)}}
\nc{\Cfour}{C_{(4)}}
\nc{\Cone}{C_{(1)}}
\nc{\Cthree}{C_{(3)}}
\nc{\Cfive}{C_{(5)}}
\nc{\Cn}{C_{(n)}}

\nc{\IGGGG}{{I_4(\cG,\cG,\cG,\cG)}}
\nc{\IGGGQ}{{I_4(\cG,\cG,\cG,\cQ)}}
\nc{\IGGQQ}{{I_4(\cG,\cG,\cQ,\cQ)}}
\nc{\IGQQQ}{{I_4(\cG,\cQ,\cQ,\cQ)}}
\nc{\IQQQQ}{{I_4(\cQ,\cQ,\cQ,\cQ)}}

\nc{\IpGGG}{{I_4(\cG,\cG,\cG)}}
\nc{\IpGGQ}{{I_4(\cG,\cG,\cQ)}}
\nc{\IpGQQ}{{I_4(\cG,\cQ,\cQ)}}
\nc{\IpQQQ}{{I_4(\cQ,\cQ,\cQ)}}

\nc{\IGQ}{\langle \cG,\cQ \rangle}

\nc{\IGGGQQQ}{\langle \IpGGG,\IpQQQ \rangle}

\nc{\equ}{{\rm eq}}
\def\Im{{\rm Im \hspace{0.5mm} }}

\def\Re{{\rm Re \hspace{0.5mm}}}

\nc{\vol}{{\rm vol}}
\nc{\Ainf}{A_{\infty}}
\nc{\End}{{\rm End}}
\nc{\Ext}{{\rm Ext}}
\nc{\IIB}{{\rm IIB}}
\nc{\Ad}{{\rm Ad}}
\nc{\IIA}{{\rm IIA}}
\nc{\AdS}{{\rm AdS}}
\nc{\CFT}{{\rm CFT}}
\nc{\diag}{{\rm diag}}
\nc{\Log}{{\rm Log}}
\nc{\Dslash}{\ensuremath \raisebox{0.025cm}{\slash}\hspace{-0.32cm} D}
\nc{\cDslash}{\ensuremath \raisebox{0.025cm}{\slash}\hspace{-0.32cm} \cD}
\nc{\omslash}{\om\!\!\!/}
\nc{\no}{\!:\!\!}
\nc{\ointdz}{\oint\frac{dz}{2\pi i}}
\nc{\ointdzone}{\oint\frac{dz_1}{2\pi i}}
\nc{\ointdztwo}{\oint\frac{dz_2}{2\pi i}}
\nc{\ointdzb}{\oint\frac{d\zbar}{2\pi i}}
\nc{\ointdzbone}{\oint\frac{d\zbar_1}{2\pi i}}
\nc{\ointdzbtwo}{\oint\frac{d\zbar_2}{2\pi i}}
\nc{\dz}{\frac{dz}{2\pi i}}
\nc{\dzb}{\frac{d\zbar}{2\pi i}}
\nc{\bpm}{\begin{pmatrix}}
\nc{\epm}{\end{pmatrix}}
 \nc{\bitem}{\begin{itemize}}
 \nc{\eitem}{\end{itemize}}
 \nc{\exercise}{\vskip 2mm \noindent {\bf Exercise:}}
 \nc{\definition}{\vskip 2mm \noindent {\bf Definition:}}
 
\begin{document}

\vspace{0.5cm}
\begin{center}
\baselineskip=13pt {\LARGE \bf{Quarter-BPS Black Holes in AdS$_4$-NUT \\
from $\N=2$ Gauged Supergravity}}
 \vskip1.5cm 
Harold Erbin and Nick Halmagyi \\ 
\vskip0.5cm
\textit{Sorbonne Universit\'es, UPMC Paris 06,  \\ 
UMR 7589, LPTHE, 75005, Paris, France \\
and \\
CNRS, UMR 7589, LPTHE, 75005, Paris, France}\\
\vskip0.5cm
erbin@lpthe.jussieu.fr \\ 
halmagyi@lpthe.jussieu.fr \\ 

\end{center}

\begin{abstract}
We study $\cN=2$ gauged supergravity with $U(1)$ gauge group coupled to $n_v$ vector multiplets and find quite general analytic solutions for quarter-BPS black holes with mass, NUT and dyonic Maxwell charges. The solutions we find have running scalar fields and flow in the IR region to a horizon geometry of the form AdS$_2\times \Sig_g$.
\end{abstract}
\newpage
\tableofcontents

\section{Introduction}

Black holes in AdS space have been studied in great detail within the context of holography \cite{Maldacena:1997re}. The supersymmetric black holes provide a laboratory where we can hope to examine the statistical underpinnings of the Bekenstein-Hawking entropy \cite{Strominger:1996sh}. In this work we continue the study of such objects within the context of $\cN=2$ gauged supergravity in four dimensions by considering static AdS$_4$ black holes with a non-vanishing NUT charge and where the scalar fields take non-trivial profiles.

The canonical example of AdS$_4$-NUT black holes can be found from a limit of the Plebanski-Demianski solution \cite{Plebanski:1976gy} and its supersymmetric structure has been studied by embedding this family of solutions into minimal gauged supergravity \cite{AlonsoAlberca:2000cs,Caldarelli2003, Martelli:2012sz}. In these solutions the scalar fields are constant and obtaining generalizations to include analytic solutions for non-constant scalar fields has proved to be somewhat non-trivial. 

For black holes with constant scalars, there are two branches of BPS solutions, preserving one-quarter and one-half of the supersymmetry \cite{AlonsoAlberca:2000cs}. More precisely they preserve two and four real supercharges respectively.  When the horizon has positive curvature, i.e. $\Sig_g=S^2$, this has been shown to agree with preservation of the supersymmetry on the boundary theory \cite{Martelli:2012sz}. 

When the NUT charge vanishes, the half-BPS solutions are electrically charged and have naked singularities while the quarter-BPS solutions can be regular if the horizon is taken to have constant negative curvature, more precisely $\Sig_g=\HH^2/\Gam$ for some discrete group $\Gam\subset SL(2,\RR)$. The half-BPS solutions with vanishing NUT charge admit a generalization to allow for running scalars which are also purely electrically charged\footnote{When loosely discussing this previous work we refer to electric and magnetic charges with respect to the symplectic frame where the STU-model described in appendix \ref{app:STU} has prepotential $F=-2i\sqrt{X^0X^1X^2X^3}$. We will be much more precise in the bulk of the text.} with naked singularities and are known as the {\it superstar} geometries \cite{Sabra:1999ux}. However, the quarter-BPS solutions can be generalized to include running scalars where regular solutions can be obtained, the first such analytic solutions were purely magnetically charged and were found in the STU-model by Cacciatori and Klemm\footnote{see \cite{Hristov:2010ri, Dall'Agata:2010gj} for further analysis of these solutions.} \cite{Cacciatori:2009iz}.  They can be regular for all values of the horizon curvature $\kappa$ and this was generalized to arbitrary symmetric very special K\"ahler manifolds in \cite{Gnecchi:2013mta}. In \cite{Halmagyi:2013uza, Katmadas:2014faa} solutions with particular electric and magnetic charges were shown to be obtained by symmetry from the solutions  of \cite{Cacciatori:2009iz, Gnecchi:2013mta}. Finally the analytic quarter-BPS solution with arbitrary electric and magnetic charges was obtained in \cite{Halmagyi:2014qza}. These quarter BPS solutions have admitted generalizations to the  non-BPS sector in \cite{Toldo:2012ec, Gnecchi:2012kb, Hristov:2013sya, Gnecchi:2014cqa} and in fact a very general solution for static non-BPS AdS$_4$ black holes in the STU-model was found in \cite{Chow:2013gba}. It is currently not understood how the solutions of \cite{Chow:2013gba} are related to the known quarter-BPS solutions; since AdS black holes allow for scalar hair and thus fixing the electromagnetic charges does not uniquely specify a solution, it is not known whether the BPS limit of the solutions in \cite{Chow:2013gba} are equivalent to the black holes of \cite{Cacciatori:2009iz, Halmagyi:2013uza, Halmagyi:2014qza}.

With non-vanishing NUT charge, one particular family of quarter-BPS black holes with running scalars and purely magnetic charges has been found in the $F=-X^0X^1$ model \cite{Colleoni:2012jq}. In the current work we present a very  general analytic solution for Fayet-Iliopoulos (FI) gauged supergravity theories where the scalar manifold $\cM_v$ is a symmetric very special K\"ahler manifold. Our ansatz allows for electric and magnetic charges which are then constrained only by the supersymmetry conditions. Our methods are a continuation of those employed in \cite{Halmagyi:2014qza, Halmagyi:2013qoa} and the solutions we find also have IR regions which are AdS$_2\times \Sig_g$ for an arbitrary genus Riemann surface. When the NUT charge vanishes, all BPS black holes have such an IR region but with non-zero NUT charge the constant scalar black hole can have more general IR boundary conditions. It is plausible that there are supersymmetric AdS$_4$-NUT solutions with running scalars and IR regions different from AdS$_2\times \Sig_g$ but these would most likely be horizon-free, we do not find such solutions in the current work. 

From purely local considerations, the addition of NUT charge is quite natural since it preserves an $SU(2)$ symmetry. Nonetheless one should recall that AdS-NUT is plagued by closed timelike curves unless $\kappa=-1$ (see \cite{Astefanesei:2004kn} for a recent discussion), nonetheless there has recently been some interesting work on understanding AdS-NUT from the dual fluid \cite{Caldarelli:2012cm, Leigh:2011au, Leigh:2012jv}. Another useful avenue to make use of the bulk NUT charge is to continue the solutions to Euclidean signature and compare with localization computations in the dual three dimensional CFT. Indeed this is our main motivation for the current work, namely to study these black holes holographically through an embedding into M-theory where the dual CFT is known. When $\cM_v =\bslb SU(1,1)/U(1)\bsrb^3$ and for a particular choice of gauging parameters, this theory is known to be a truncation of eleven-dimensional supergravity on $S^7$. For such models, our black holes correspond to the holographic dual of the ABJM theory \cite{Aharony:2008ug} on curved manifolds $M_3$. We work in Lorentzian signature but the Euclidean continuation contains solutions dual to ABJM on Seifert spaces (given by a $U(1)$ bundle over $\Sig_g$), including the Lens spaces $S^3/\ZZ_N$, where supersymmetry has been preserved by twisting the theory with respect to a general  $U(1)\subset SU(4)_{\cR}\times U(1)_{\cR}$. From an $\cN=2$ point of view this includes flavour as well as $\cR$-symmetries. We will return to this holographic study in a future publication.

This paper is organized as follows. In section \ref{constantscalar} we revisit the constant scalar black hole in AdS$_4$-NUT and analyze the root structure of $g_{tt}$ for the two BPS branches of solutions. In section \ref{Symplectic} we introduce the ansatz and the form of the BPS equations which we will utilize. These BPS equations are derived in appendix \ref{app:BPSDerivation}. In section \ref{sec:Roots} we will discuss some generalities about the ansatz emphasizing the root structure of the $g_{tt}$ component of the metric. In section \ref{sec:PairDoubleRoot} we present the basic class of our solutions which are then generalized in section \ref{sec:DoubleRoot}. In section \ref{sec:Independent} we present the well known constant scalar solution in our notations. Most of the calculations are relegated to the appendices.

\section{Motivation From the Constant Scalar Black Hole}\label{constantscalar}

Before commencing the analysis of AdS$_4$ black holes with NUT charge and running scalars, we revisit the AdS$_4$-NUT black holes in Einstein-Maxwell theory. The metric is given by
\bea
ds^2 &=& \frac{e^{2V}}{r^2+N^2}(dt+2N \tf(\tha) d\phi)^2 -  \frac{r^2+N^2}{e^{2V}} dr^2 - (r^2+N^2) \bslb d\tha^2 + f^2(\tha) d\phi^2 \bsrb  \\
e^{2V}&=& g^2  (r^2+N^2)^2 +(\kappa + 4 g^2 N^2) (r^2-N^2)-2Mr +P^2 +Q^2 \label{MINe2V}
\eea
where the familiar three possible horizon geometries $\Sig_g$ are considered\footnote{Our solutions admit an $SL(2,\RR)$ symmetry which acts on the horizon co-ordinates $(\tha,\phi)$ and when $N\neq0$ on the time direction $t$, thus one can quotient $\RR^2,\HH^2$ by a discrete subgroup to obtain a compact horizon.}:
\be
f(\tha)=\left\{ \barr{ll} \sin \tha& S^2\quad(\kappa=1) \\ 1 & \RR^2\quad (\kappa=0) \\ \sinh \tha & \HH^2\quad (\kappa=-1) \earr \right.\, \quad {\rm and}\quad
\tf(\tha)=\left\{ \barr{ll} f'(\tha)& S^2\quad(\kappa=1) \\ \tha & \RR^2\quad (\kappa=0) \\ f'(\tha) & \HH^2\quad (\kappa=-1) \earr \right. \,.\label{Horizons}
\ee
The gauge field is given by
\bea
A_t&=& \frac{Qr - NP }{r^2+N^2}\,,\qquad A_\phi= \frac{f\bslb  P(r^2-N^2)+2NQr\bsrb }{r^2+N^2}\,.
\eea

For $\kappa=0,1$ AdS-NUT and thus also asymptotically AdS-NUT solutions have closed time-like curves in addition to the Dirac-Misner string \cite{Misner:1963fr}. For $\kappa=-1$ there is a claim \cite{Astefanesei:2004kn} that there exists a upper bound on $N$, below which the solution will be free of closed time-like curves. We proceed nonetheless with a view towards ultimately continuing our solutions to Euclidean space. In this section we analyze the root structure of \eq{MINe2V} in order to understand the possible supersymmetric horizons we should expect in our solutions with running scalars in the following sections. In particular we compute the conditions for $e^{2V}$ to have real roots while also preserving supersymmetry.

Romans initiated the study of the supersymmetry properties of these asymptotically AdS solutions \cite{Romans:1991nq} with vanishing NUT charge by using the canonical embedding of Einstein-Maxwell theory into minimal gauged supergravity \cite{Freedman:1976aw}.  He found two classes of BPS solutions, preserving one half or one quarter supersymmetry:
\be
\barr{lll}
&{\underline{\rm Vanishing\ NUT\ Charge}\ \cite{Romans:1991nq}}& \\
{\rm \frac{1}{4}-BPS}:&\quad  M=0  & \quad P= \pm \frac{1}{2g} \\
{\rm \frac{1}{2}-BPS}:& \quad M= |Q| &\quad P= 0 \\
\earr
\ee
The half BPS solution has a naked singularity for any $\kappa$  and is sometimes referred to as a superstar \cite{Myers:2001aq}.  The quarter BPS solution has a naked singularity in general but for $\kappa=-1$ and $Q=0$ it has an extremal horizon of the form AdS$_2\times \HH^2$ \cite{Caldarelli1999} and thus may be called a black hole. 

This was generalized to include NUT charge and arbitrary $\kappa$ in \cite{AlonsoAlberca:2000cs} where the two classes of BPS black holes were found to satisfy the following relations amongst their parameters:
\be
\barr{lll}
&{\underline{\rm Non-Vanishing\ NUT\ Charge}\ \cite{AlonsoAlberca:2000cs}}& \\
{\rm \frac{1}{4}-BPS}:&\quad  M= |2g N Q|  & \quad P= \pm \frac{\kappa  + 4 g^2 N^2}{2g} \\
{\rm \frac{1}{2}-BPS}:& \quad M= |Q\sqrt{\kappa + 4g^2 N^2}|  &\quad P= \pm N \sqrt{\kappa + 4g^2 N^2} \\
\earr
\ee
with both $(N,Q)$ unconstrained. 

The root structure of $e^{2V}$ simplifies on these two BPS solution branches. With 
\be
e^{2V}=g^2 (r-r_1^+)(r-r_1^-)(r-r_2^+)(r-r_2^-)
\ee
we find
\bea
{\rm \frac{1}{4}-BPS} && \left\{
\barr{l}
r_1^{\pm}= -i \Bslb N \pm \frac{1}{\sqrt{2} \, g}\sqrt{4g^2 N^2 - 2igQ + \kappa} \Bsrb \\
r_2^{\pm}= i \Bslb N \pm \frac{1}{\sqrt{2} \, g}\sqrt{4g^2 N^2 + 2igQ + \kappa} \Bsrb \\
\earr\right. \\
& &  \non \\
&&  \non \\
{\rm \frac{1}{2}-BPS} &&\left\{
\barr{l}
r_1^{\pm}= -\frac{i}{2g}\Bslb \sqrt{\kappa + 4 g^2 N^2} \pm  \sqrt{8 g^2 N^2 -4 i g Q +\kappa}  \Bsrb \\
r_2^\pm = \frac{i}{2g}\Bslb \sqrt{\kappa + 4 g^2 N^2} \pm \sqrt{8 g^2 N^2 + 4 i g Q +\kappa}\Bsrb \\
\earr\right.
\eea
For the quarter-BPS branch, a real root of $e^{2V}$ requires
\bea
Q^2=-2N^2 (2g^2 N^2 + \kappa)
\eea
which clearly implies $\kappa=-1$. Then the quarter-BPS branch of solutions has an extremal horizon at
\be
r_1^-=r_2^-=\frac{\sqrt{1-\kappa -4g^2 N^2}}{2\sqrt{2}\,g}\,.
\ee
which for is manifestly positive. Interestingly, according to the criteria of \cite{Astefanesei:2004kn} solutions with $\kappa=-1$ and $0\leq 2g^2N^2\leq1$ are free from closed time-like curves. 

For the half-BPS branch, reality of the magnetic charge requires $\kappa + 4 g^2 N^2>0$ but then a real zero of $e^{2V}$ requires
\be
Q^2=-N^2 (4g^2 N^2 +\kappa)
\ee
which is a contradiction. We conclude that for the half BPS branch with $N\neq0$, space-time continues through $r=0$ where nothing shrinks, to negative $r$. This may be contrasted with the analysis in Euclidean space in \cite{Martelli:2012sz} where the Euclidean solutions have single roots and thus bolts in the interior. The quantitative difference in the root structure of Lorentzian and Euclidean solutions is due to the analytic continuation of $N$ when passing between the two signatures.

In the next sections we will find analytic solutions for AdS$_4$-NUT black holes with running scalar fields and by construction these have double roots in $e^{2V}$. This seems to be a physically reasonable construction since a black hole with a single root of $e^{2V}$ would have a finite temperature, we expect supersymmetric black holes to be extremal. In principle there should exist solutions to the black hole ansatz we will use which have no horizons at all (no real roots of $e^{2V}$), these will prove to be beyond our analysis. We restrict our analysis to quarter-BPS solutions, leaving the analysis of half-BPS solutions to future work.

\section{Symplectic Covariant BPS Equations} \label{Symplectic}

In this section we present the ansatz and supersymmetry equations for the AdS$_4$-NUT black holes we consider. In general, the NUT charge provides a source for the Maxwell fields, with some effort we can recast the supersymmetry equations in terms of the conserved Maxwell charges. While the derivation of the supersymmetry equations is presented in appendix \ref{app:BPSDerivation}, we have tried to present enough material here to orient the reader and set up notation, the key result is the final form of the equations \eq{BPSEqs} which is valid when $\cM_v$ is a symmetric very special K\"ahler manifold. The symplectic covariant framework we employ is heavily influenced by \cite{Dall'Agata:2010gj}, we have found that solving for general models in this formalism is more tractable than choosing a particular model in the formalism with just electric gaugings.

\subsection{The Black Hole Ansatz}

In this subsection we describe the ansatz we use for the metric, gauge fields and scalars. The metric is 
\bea
ds^2 &=& e^{2U}(dt+2N \tf(\tha) d\phi)^2 - e^{-2U} dr^2 - e^{2(V-U)} \bslb d\tha^2 + f^2(\tha) d\phi^2 \bsrb 
\eea
where the possible horizon geometries $\Sig_g$ we consider are the same as in \eq{Horizons}\footnote{The case of $\kappa=0$ is somewhat independent, we will work through the text with $\kappa=\pm1$ but one can easily check that our final equations \eq{FinalEq1}-\eq{FinalEq7} are valid for $\kappa=0$ as well. We will be somewhat more precise at the end of appendix \ref{app:BPSDerivation}.}. 

The gauge field is given by 
\bea
A^\Lam &=&   \tq^\Lam dt -\kappa P^\Lam f' d\phi \quad \Ra \quad
F^\Lam=  \tq'^\Lam dr\w (dt+2N f' d\phi )+   P^\Lam f d\tha \w d\phi  \,,\quad \Lam=0,\ldots , n_v \non
\eea
and the complex scalar fields depend only on the radial co-ordinate $r$:
\be
z^i=z^i(r)\,,\qquad i=1,\ldots ,n_v\,.
\ee
As is quite standard in $\cN=2$ supergravity, we package the scalar fields into sections of an $Sp(2n_v+2,\RR)$-bundle over the scalar manifold $\cM_v$
\be
\cV=\bpm L^\Lam \\ M_\Lam \epm  = e^{K/2} \bpm X^\Lam \\ F_\Lam \epm
\ee
where $K(z^i,\zbar^{\ibar})$ is the K\"ahler potential on $\cM_v$. For the black hole solutions we study, the expressions for $\cV$ are simpler than for the scalar fields themselves.  The dual gauge field strength is given by
\be
G_\Lam \equiv  \cR_{\Lam\Sig} F^\Sig - \cI_{\Lam\Sig} *F^{\Sig}\,,
\ee
we then make a symplectic vector out of the gauge fields as follows:
\be
\hcQ= \bpm P^\Lam \\ Q_\Lam \epm\,,\qquad Q_\Lam =- \cI_{\Lam\Sig}\tq'^\Sig e^{2(V-U)} +P^\Sig \cR_{\Lam\Sig} \,.
\ee
As we will see in the next subsection, $\hcQ$ does not contain the conserved Maxwell charges but is sourced by the NUT charge.

The supergravity theories we consider are also specified by a symplectic vector of gauging parameters
\be
\cG= \bpm g^\Lam \\ g_\Lam \epm \,.
\ee
One can always find a symplectic frame where the gaugings are all electric $(g^\Lam=0)$ and thus the gravitino is minimally coupled to electric gauge fields but we will work in a formalism which preserves the symplectic covariance and thus we allow for both electric and magnetic gauging parameters. Note that the sections $\cV$ are not invariant under symplectic transformations so if one would insist on having electric gauge couplings, one would need to allow for a much more general form of the sections. Exactly as in \cite{Halmagyi:2013qoa, Halmagyi:2014qza}, in this work our strategy is  to consider sections $\cV$ which come from a cubic prepotential and allow for dyonic gauge couplings.

\subsection{The Supersymmetry Equations}\label{sec:BPSEqs}

In this subsection we present the form of the supersymmetry equations which we will solve. This involves specializing $\cM_v$ to be a symmetric very special K\"ahler manifold\footnote{This means that $\cM_v$ is a special K\"ahler manifold with cubic prepotential $F=-d_{ijk}\frac{X^iX^jX^k}{X^0}$ and in addition is a symmetric space.}.

For the quarter-BPS solutions we study, the supersymmetry parameter is given by 
\be
\eps_A= e^{(U+i\psi)/2} \eps_{0A}
\ee
where $\eps_{0A}$ is an $SU(2)$ doublet of constant spinors and the following two projectors are enforced:
 \bea
 \eps_{0A}&=&i  \eps_{AB}\gam^{0}  \eps_0^B\,,  \\
\eps_{0A}&=&    -   (\sigma^3)_A^{\ B} \gam^{01} \epsilon_{0B} \,.
\eea
Both $(U,\psi)$ are functions of the radial co-ordinate only and thus the spinors $\eps_A$ are independent of the co-ordinates on the horizon $\Sig_g$.

The symplectic covariant form of the equations is a generalization of \cite{Dall'Agata:2010gj} to include NUT charge. Much as in \cite{Halmagyi:2014qza} where the NUT charge was zero, we find that this form of the BPS equations leads fairly directly to a solution. There are two symplectic vectors worth of equations
\bea
2 e^{2V} \del_r \Bslb \Im\blp e^{-i\psi}e^{-U}\cV \brp \Bsrb &=& \Bslb  8 e^{2(V-U)} \Re\blp e^{-i\psi} \cL\brp + 4N \kappa e^{U} \Bsrb\Re \blp e^{-i\psi}\cV \brp + \hcQ - e^{2(V-U)} \Om \cM \cG \non \\ && \label{ImVBPS}\\
2\del_r \bslb\Re ( e^Ue^{-i\psi}\cV)\bsrb&=& e^{2(U-V)} \Om \cM \hcQ + \cG \label{ReVBPS}
\eea
but since $\Re \cV$ and $\Im \cV$ are not independent, neither are \eq{ImVBPS}-\eq{ReVBPS}. Indeed \eq{ImVBPS} has $2n_v+2$ real components from which one can extract equations for the $n_v$ complex scalar fields, the spinor phase $\psi$ and the metric function $U$. Nonetheless the form of \eq{ReVBPS} is quite useful as we will see below.

There is an equation for the metric function $V$
\bea
\del_r \blp e^V \brp &=& 2  e^{V-U} \Im \blp  e^{-i\psi} \cL\brp  \label{VBPSEq}
\eea
and the symplectic covariant form of Maxwell's equations is
\bea
\del_r \hcQ&=& -2N\kappa e^{2(U-V)} \Om \cM \hcQ  \,. \label{MaxEqhcQ}
\eea
In addition one must impose the single real constraint
\be
0=N\kappa e^{3U-2V} + \Re(e^{-i\psi} \cL) + e^{2(U-V)}\Im ( e^{-i\psi}\hcZ) \label{BPSConstraint}
\ee
where 
\be
\hcZ=\langle \hcQ,\cV \rangle\,, \qquad \cL=\langle \cG,\cV \rangle
\ee
as well as the BPS-Dirac quantization condition
\be
\langle \cG , \hcQ+4N\kappa e^U \Re (e^{-i\psi}\cV )\rangle =  \kappa \,.
\ee
From \eq{ImVBPS} and \eq{BPSConstraint} one could derive the following form of the differential equation for the phase
\be
\psi'+A_r=-2  e^{-U}\Re (e^{-i\psi} \cL) - N\kappa e^{2(U-V)} 
\ee
but this is not an independent equation and we will not use this further in our analysis.

When the NUT charge is non-zero, $N\neq 0$,  then using Maxwell's equation \eq{MaxEqhcQ} we can integrate \eq{ReVBPS} to find
\be
2\Re ( e^Ue^{-i\psi}\cV)=-\frac{1}{2N\kappa } \hcQ+\cG r - \frac{1}{2N\kappa} \cQ \label{ReVQ}
\ee
where we have introduced a symplectic vector of integration constants 
\be
\cQ=\bpm p^\Lam \\ q_\Lam \epm
\ee 
which we recognize as the conserved Maxwell charges. The Dirac quantization condition then becomes independent of the NUT charge:
\be
\langle \cG , \cQ\rangle =  -\kappa \label{DiracQhat}
\ee
and we can transform \eq{ImVBPS} into an equation which depends on the constants $\cQ$ instead of $\hcQ$:
\bea
2 e^{2V} \del_r \Bslb \Im\blp e^{-i\psi}e^{-U}\cV \brp \Bsrb &=&   8  e^{2(V-U)} \Re\blp e^{-i\psi} \cL\brp \Re \blp e^{-i\psi}\cV \brp  +2N \cG r -\cQ-e^{2(V-U)} \Om \cM \cG\,.\non\\
&&  \label{VimEq1}
\eea

\subsection{Symmetric Scalar Manifolds}\label{sec:Symmetric}

We find that \eq{VimEq1} is not yet an optimal form of the equations since it involves $\Re\cV$ as well as $\Im \cV$, not to mention the somewhat complicated field-dependent matrix $\cM$.  Recall that $\cV$ has $2n_v+2$ complex components but there are only $n_v$ complex fields $z^i$,  so we do not need to solve individually for both $\Re \cV$ and $\Im \cV$.
When $\cM_v$ is symmetric we can make significant further progress on these equations, namely we can use an identity for any symplectic vector $A$  \cite{Halmagyi:2014qza, Katmadas:2014faa} :
\be
I'_4(A,\Im \cV,\Im\cV)=4  \langle A,\Im\cV \rangle \Im \cV+8   \langle A,\Re\cV \rangle \Re \cV -\Om \cM A
\ee
and find that \eq{VimEq1} becomes
\be
2 e^{V} \del_r \blp \Im \tcV  \brp =I'_4(\cG,\Im \tcV,\Im\tcV)  +2N \kappa \cG r - \cQ
\label{BPSnice}\ee
where we have introduced a complex rescaling of the sections
\be
\tcV=e^{-i\psi}e^{V-U} \cV \,.
\ee
Happily \eq{BPSnice} now depends only on $\Im \tcV$ and not $\Re \tcV$

The constraint \eq{BPSConstraint} requires some further manipulation to eliminate the dependence on $\Re \tcV$ and to express it in terms of $\cQ$ as opposed to $\hcQ$. 
Using \eq{ReVQ},\eq{VimEq1} as well as the identity 
\bea
\Re \tcV &=& -\frac{1}{3} e^{2(U-V)} I_4'(\tcV,\tcV,\tcV) 
\eea
and the contraction of \eq{BPSnice} with $\Im\tcV$ 
\bea
2 e^{V}\langle \Im \tcV,  \del_r \Im \tcV \rangle   &=& I'_4(\cG,\Im \tcV,\Im \tcV,\Im\tcV)  -2N \kappa r  \langle \cG ,\Im \tcV\rangle +\langle  \cQ,\Im \tcV\rangle\,,
\eea
we find that \eq{BPSConstraint} can be re-written as
\bea
0&=&  3N\kappa e^{V} + 2 e^{V}\langle \Im\tcV,\del_r(\Im \tcV)\rangle - 4N\kappa r \langle \cG,\Im\tcV \rangle +  2 \langle \cQ,\Im\tcV\rangle  \label{constraint2}\,.
\eea

In summary, at this point in the analysis we must solve \eq{VBPSEq},\eq{BPSnice} and \eq{constraint2} for $\tcV$ and $e^V$, from this we can extract the solution for the metric, scalar fields, gauge fields and supersymmetry parameter. On this solution space we then must impose the BPS-Dirac quantization condition which will be important for regularity of any given solution.
For convenience we summarize here our final form of the BPS equations:
\be \fbox{$\barr{rcl} 
2 e^{V} \del_r \blp \Im \tcV  \brp &=&I'_4(\cG,\Im \tcV,\Im\tcV)  +2N \kappa \cG r - \cQ \\
\del_r \blp e^V \brp &=& 2    \langle \cG,\Im\tcV \rangle  \\
 3N\kappa e^{V} &=&  -2 e^{V}\langle \Im\tcV,\del_r(\Im \tcV)\rangle+4N\kappa r \langle \cG,\Im\tcV \rangle -  2 \langle \cQ,\Im\tcV\rangle  \\
\langle \cG , \cQ\rangle &=&  -\kappa 
 \earr$}\label{BPSEqs}
\ee
We repeat here that these are valid for $\kappa=\pm1$, to get the equations for $\kappa=0$ one must send $N\kappa\ra N$ and then $\kappa\ra 0$.

\section{Generalities Regarding the IR Geometry}\label{sec:Roots}

The central point of the ansatz we use to solve \eq{BPSEqs} is that, motivated by the constant scalar solution, $e^{2V}$ is taken to be a quartic polynomial in $r$:
\be
e^{2V}=\sum_{i=0}^4 v_i r^i\,.
\ee
The various possible branches of solutions could then be classified by their root structure:
\be
\barr{rl}
{\rm triple\ root}:& v_0=v_1=v_2=0 \\
{\rm pair\ of\ double\ roots}:& v_0=v_1=0\,,\quad v_3= 2\sqrt{v_2v_4}\\
{\rm single\ double\ root}:& v_0=v_1=0 \\ 
{\rm two\ pairs\ of\ conjugate\ roots}: & v_3=0 \\
{\rm at\  least\ two\ real\ roots}:& v_0=0
\earr
\ee

When $e^{2V}$ has a real double root, we can move this root to zero through a shift in the radial co-ordinate, this sets $v_0=v_1=0$. Having a second double root then sets $v_3= 2\sqrt{v_2v_4}$. The IR behavior around $r=0$ is then an exact AdS$_2\times \Sig_g$ geometry with radii $(R_1,R_2)$ respectively. The metric functions evaluate at the horizon to
\be
e^{2V}\sim r^2 v_2\,,\quad e^{2(V-U)}\sim R_2^2 \quad \Ra\quad v_2= \frac{R_2}{R_1}\,.
\ee
We note that a metric of the form
\be
ds_4^2= -\frac{r^2}{R_1^2} (dt+2N\cos\tha d\phi)^2 +\frac{R_1^2 dr^2}{r^2}+ R_2^2 d\Sig_g^2
\ee
approaches AdS$_2\times \Sig_g$ in the limit 
\be
r\ra \eps r \,,\quad t\ra t/\eps\,,\quad \eps\ra 0\,.
\ee

With vanishing NUT charge, the solutions of \cite{Cacciatori:2009iz, Gnecchi:2013mta} have a pair of double roots and in the symplectic frame where the gauge couplings are electric, the charges are purely magnetic. We will find below, the generalization of these solutions to include NUT charge but with $N\neq 0$ a pair of double roots allows for dyonic charges.

The solutions of \cite{Halmagyi:2014qza} have just a single double root and this appears to exhaust the possibilities when the NUT charge vanishes. We will also find the generalization of solutions with this root structure to include NUT charge. 

As reviewed in section \ref{constantscalar}, the constant scalar solution with electric-magnetic charges and NUT charge has a quarter-BPS branch where there are no real roots. For such solutions, one cannot use the shift freedom in $r$ to set $v_0=0$ but instead one can set $v_3=0$. We will find this constant scalar solution with no real roots within our ansatz but its generalizations to running scalars will prove to be beyond our reach, that is not to say they do not exist. One can also quite easily rule out the possibility of a quarter-BPS black hole with a triple root, although as found in \cite{Romans:1991nq} there are such non-BPS examples (referred to there as {\it ultracold} black holes).

\section{Pair of Double Roots}\label{sec:PairDoubleRoot}

When there is a pair of double roots our ansatz is:
\bea
e^{2V}&=& r^2 \Blp v_4  r^2+2\sqrt{v_2v_4} r +v_2\Brp\,, \\
\Im \tcV&=& \frac{1}{\eps\sqrt{2\langle \cG, A_1 \rangle}}A_1 + r A_3
\eea
where $(A_1,A_3)$ are symplectic vectors which we must determine and we include a sign $\eps=\pm1$ to keep track of both branches of the square root. The IR and UV asymptotics completely fix the solution, the BPS equations then over-constrain this ansatz and for a solution to exist there must be significant cancellation. We have introduced this particular normalization of $A_1$ to make contact with expressions elsewhere.

We first solve the second equation of \eq{BPSEqs} to get
\be
\sqrt{v_2}=\eps \sqrt{2 \langle  \cG,A_1\rangle}\,,\qquad \sqrt{v_4} = \langle \cG,A_3 \rangle\,.
\ee
and then expand the BPS equations \eq{BPSnice}  in $r$ to get:
\bea
0&=& I'_4(\cG,A_3,A_3)-2 \langle \cG,A_3\rangle A_3 \label{PairBPSEq1}\\
0&=& I'_4(\cG,A_1,A_3) - 2 \langle \cG,A_1\rangle A_3 + N \kappa \eps \sqrt{2 \langle \cG,A_1 \rangle} \,\cG \label{PairBPSEq2} \\
0&=& I'_4(\cG,A_1,A_1)-2\langle \cG,A_1 \rangle \cQ \,. \label{PairBPSEq3}
\eea
The constraint \eq{constraint2} is also expanded and we get
\bea
0&=&  \sqrt{2}\langle A_1,A_3\rangle-N\kappa \eps  \langle \cG,A_1\rangle^{1/2} \label{PairConstr1}\\
0&=&\sqrt{2} N\kappa \eps \langle \cG,A_1\rangle^{3/2} +  \langle \cG,A_3\rangle \langle \cQ,A_1\rangle +2 \langle \cG,A_1\rangle \blp \langle \cQ,A_3\rangle+ \langle A_1,A_3\rangle\brp \label{PairConstr3} \\
0&=& \langle \cQ,A_1 \rangle\label{PairConstr4}
\eea

All the free parameters are fixed by the UV and IR asymptotics. From the UV we get
\footnote{We will economize a little and at times use $I_4(\cG)$ and $I'_4(\cG)$ instead of $\IGGGG$ and $I'_4(\cG,\cG,\cG)$. See appendix \ref{app:quartic} for the conversion factors}
\bea
A_3&=& \frac{I_4'(\cG)}{4 I_4(\cG)^{1/4}}\,,\qquad v_4= \sqrt{I_4(\cG)}\,,
\eea
where we have appealed to \cite{Halmagyi:2013qoa} to fix the normalization of $A_3$.
The solution for $A_1$, found from the IR equation \eq{PairBPSEq3}, is the same as in \cite{Halmagyi:2014qza}. In deriving this we have made use of various identities for the quartic invariant which can be found in appendix \ref{app:identities}:
\be
A_1= a_1 I'_4(\cG,\cG,\cG)+ a_2 I'_4(\cG,\cG,\cQ)+ a_3 I'_4(\cG,\cQ,\cQ)+ a_4 I'_4(\cQ,\cQ,\cQ)
\ee
with
\bea
a_1&=& -\frac{a_3 \IGQQQ}{3\IGGGQ} \label{PairHorizon1}\\
a_2&=& \frac{a_3}{6}\frac{ \IGGGQ \IGQQQ^2}{\IGGGQ^2 I_4(\cQ)-I_4(\cG) \IGQQQ^2}  \label{PairHorizon2} \\
a_3&=& \frac{9\blp \IGQQQ I_4(\cG)-\IGGGQ I_4(\cQ)  \brp }
{\IGQQQ \IGQQQ \blp \langle I'_4(\cG,\cG,\cG), I'_4(\cQ,\cQ,\cQ)\rangle  +\kappa  \IGGQQ\brp }  \label{PairHorizon3}\\
a_4&=& - \frac{a_2 \IGGGQ}{3 \IGQQQ} \,. \label{PairHorizon4}
\eea
In particular one should note that in deriving \eq{PairHorizon1}-\eq{PairHorizon4}, as a computational tool we have enforced the constraint 
\bea
0&=&4\IGGGQ^2 I_4(\cQ)+ 4\IGQQQ^2 I_4(\cG) -\IGGGQ\IGGQQ\IGQQQ \non \\
&& \label{AdS2Constraint}
\eea
by solving it for $\IGGQQ$. This constraint was first derived in \cite{Halmagyi:2013qoa} as a BPS condition for quarter-BPS AdS$_2\times \Sig_g$ vacua which we find here in the IR. Since the constraint is independent of the radius, it should be imposed on the whole black hole solution.

The effect of the NUT charge is through \eq{PairBPSEq2} as well as the constraints \eq{PairConstr1} and \eq{PairConstr3}. We find that these three equations are redundant and there is a single non-trivial constraint on the system:
\bea
N\kappa\eps &=& -\frac{\IGGGQ^{2} \IGQQQ}{144 \sqrt{2} \, I_4(\cG)^{1/4}} \times \non \\
&& \hspace{-1.5cm} \frac{\sqrt{18\langle \cG,\cQ \rangle \IGGQQ-\langle I'_4(\cQ,\cQ,\cQ),I'_4(\cG,\cG,\cG) \rangle } }{\sqrt{\blp I_4(\cG)\IGQQQ^2-\IGGGQ^2I_4(\cQ) \brp^2 +16 \IGGGQ^3\IGQQQ^3}} \,.\label{NConstraintDoubleRoot}
\eea

When $N=0$ then \eq{NConstraintDoubleRoot} is solved by $\IGQQQ=\IGGGQ=0$ and the solutions reduce to those in \cite{Cacciatori:2009iz, Gnecchi:2013mta, Katmadas:2014faa}.

\subsection{Example: T$^3$ Model}

We now write down a non-trivial example by restricting to the T$^3$ model\footnote{we refer to the T$^3$ model as the STU-model of appendix \ref{app:STU} with $p^1=p^2=p^3$ and $q_1=q_2=q_3$ as well as the sections $L^1=L^2=L^3$ and $M_1=M_2=M_3$.} and allowing for dyonic charges. One might first try to find the solution with the same charges $(p^1,q_0)=(0,0)$ as Cacciatori-Klemm solution \cite{Cacciatori:2009iz} but we find quite straight-forwardly that this requires $N=0$ and thus does not admit a generalization with NUT charge.

\subsubsection{$p^1=0$}
For simplicity, such that the resulting expressions are not too cumbersome, we set $p^1=0$. We can solve the constraint \eq{AdS2Constraint} with
\be
q_0= \frac{(p^0-q_1)^{3/2}}{\sqrt{2p^0}}
\ee
then the imaginary parts of the sections are given by
\bea
\Im \tcV^0&=& \frac{\eps p^0}{\sqrt{g} \sqrt{5p^0+3q_1}}+\frac{g^2 r}{\sqrt{2}}  \,,\qquad 
\Im \tcV^i= \frac{\eps \sqrt{p^0}\sqrt{p^0-q_1}}{\sqrt{8g} \sqrt{5p^0+3q_1}} \\
\Im \tcV_0&=&\frac{\eps (2p^0+q_1)\sqrt{p^0-q_1}}{\sqrt{8g}\sqrt{p^0} \sqrt{5p^0+3q_1}} \,,\qquad 
\Im \tcV_i= \frac{\eps( p^0+q_1)}{2\sqrt{g} \sqrt{5p^0+3q_1}} +\frac{g^2 r}{\sqrt{2}} \,.
\eea
and the metric components are given by
\bea
e^{2V}= r^2 \bslb 2\sqrt{2}\, g^3\, r +\eps \sqrt{g}\sqrt{5 p^0+3q_1}\bsrb^2\,.
\eea
The NUT charge is given by the relation
\be
N\kappa \eps =\frac{g^{3/2}}{2\sqrt{p^0}} \frac{(p^0-q_1)^{3/2}}{\sqrt{5p^0+3q_1}}
\ee
and the BPS Dirac quantization condition is 
\be
-\kappa= g (p^0+3q_1)\,.
\ee

When $\eps=+1$, the horizon is at $r=0$ and we find that regular solutions exist for both $\kappa=\pm1$. When $\eps=-1$ the horizon is at
\be
r=\frac{\sqrt{5p^0 + 3q_1}}{g^{5/2} \sqrt{8}}
\ee
and for the absence of zeros in $\Im \tcV$ we need
\be
g(p^0 + 3q_1 )>0
\ee
which implies $\kappa=-1$.
\subsection{Constant Scalar Solution}
One can observe the limit $p^1=q_0$ which gives the constant scalar solution. The combination of constant scalar fields and a pair of double roots in $e^V$ forces $N=0$ and as is well-known we have a hyperbolic horizon $\kappa=-1$. The solution data is given by
\bea
&& \Im \tcV^0=\Im \tcV_i =\frac{1}{2\sqrt{2}\, g} \Bslb 2g^3 r + \sqrt{p^0 g}  \Bsrb \,,\\
&& \Im \tcV_0= \Im \tcV_i= 0
\eea
and the metric components are
\bea
e^{2V}&=&2\sqrt{2}  r^2 \bslb \, g^3\, r +\sqrt{p^0 g}\bsrb^2\\
e^{2(V-U)}&=& \frac{1}{g}\bslb2g^3 r + \sqrt{p^0 g}\bsrb^2 \,.
\eea

\subsection{$F=-X^0X^1$}
We can write quite explicitly the solution when 
\be
p^0=-q_1\,,\quad p^1=q_0\,,\quad p^3=p^2\,,\quad q_3=q_2
\ee
which is equivalent to considering the prepotential
\be
F=-X^0X^1
\ee
and allowing for four arbitrary charges\footnote{Since this model is seen as a truncation from the STU model we can use our formulas for cubic prepotentials even if the prepotential is quadratic.}.
The solution to the constraint \eq{AdS2Constraint} is taken to be
\be
p^0=\frac{p^2 q_0}{q_2}
\ee
and we then find the following data:
\bea
(A_0)^0= -(A_0)_1&=& -\frac{q_2 (p^2-q_0)^2}{2((p^2)^2+q_2^2)(p^2q_0+q_2^2)(q_0^2+q_2^2)} \times \\
&& \Bslb (q_0^2-q_2^2)^2 q_2^2 +(p^2)^2(q_0^2+4q_0q_2+q_2^2)+2q_2 p^2(2 q_2^2-q_0q_2) \Bsrb  \\
 (A_0)^2= -(A_0)^3&=&  \frac{p^2(p^2-q_0)^2q_0q_2}{((p^2)^2+q_2^2)(q_0^2+q_2^2)} \\
(A_0)_0= -(A_0)^1&=& \frac{(p^2-q_0)q_2^3}{((p^2)^2+q_2^2)(q_0^2+q_2^2)} \\
(A_0)_2= -(A_0)_3&=&  -\frac{q_2 (p^2-q_0)^2}{2((p^2)^2+q_2^2)(p^2q_0+q_2^2)(q_0^2+q_2^2)} \times \\
&& \Bslb (q_0^2+q_2^2)^2 (p^2)^2 +2 p^2 q_2 (q_2^2 -2q_0^2) +q_2^2(q_0^2-4q_0 q_2 +q_2^2) \Bsrb 
\eea
The NUT charge is given by
\bea
N\kappa =-\frac{g^{3/2}}{q_2(q_2-q_0) +p^2(q_0+q_2)}
\sqrt{-\frac{((p^2)^2+q_2^2)(p^2q_0+q_2^2)(q_0^2+q_2^2)}{2q_2}} \eea
and the metric components can be obtained from 
\bea
v_2&=&(q_0-p^2) \sqrt{\frac{-2g q_2\bslb (p^2)^2 q_0^2 +4 p^2 (p^2-q_0)q_0q_2+((p^2)^2+q_2^2)q_2^2+4(p^2 -q_0)q_2^3+q_2^4 \bsrb }{(p^2)^2+q_2^2)(p^2q_0+q_2^2)(q_0^2+q_2^2)}}
\eea
and 
\be
v_4= \sqrt{8} g^3\,.
\ee

\section{Single Double Root}\label{sec:DoubleRoot}

Only a single double root is required in $e^{2V}$ in order to have an AdS$_2\times \Sig_g$ vacuum in the IR but this more general solution is somewhat more complicated. We found that in order to have a pair of double roots, there is a relation between the NUT charge and the electro-magnetic charges \eq{NConstraintDoubleRoot}, we find that there is no such constraint when requiring a single double root. The only constraint is that for AdS$_2\times \Sig_g$ vacua \eq{AdS2Constraint}.

We take the same ansatz as in \cite{Halmagyi:2014qza}:
\bea
e^{2V} &=& r^2\blp v_2 + v_3 r + v_4 r^2 \brp \\
\Im \tcV&=& e^{-V} \hA   \\
\hA&=&\Bslb  A_1 r + A_2 r^2 + A_3 r^3 \Bsrb
\eea
where $A_i$ are constant symplectic vectors whose dependence on $\cG$ and $\cQ$ we seek to determine. 
We first solve  \eq{FinalEq3} with
\bea
v_{i+1}&=&\frac{4}{i+1} \langle \cG,A_i \rangle\,,\qquad i=2,3,4\,.
\eea
The symplectic vector of BPS equations \eq{BPSnice} is then
\bea
2  e^{2V} \hA'-(e^{2V})'\hA   = I'_4(\cG,\hA,\hA)  +e^{2V}(2N  \cG r - \cQ)\label{cVEq4}
\eea
which breaks up into five components from different powers of $r$:
\bea
0&=&  I_4'(\cG, A_3,A_3)  -2 \langle \cG,A_3\rangle  A_3   \label{SectionEq1} \\
0 &=&I_4'(\cG, A_2,A_3) + N \kappa \langle \cG,A_3\rangle \cG  -2 \langle G,A_2 \rangle A_3  \label{SectionEq2}  \\
0&=&  2I_4'(\cG,A_1,A_3) +I_4'(\cG,A_2,A_2) -8  \langle \cG,A_1 \rangle A_3 - \langle \cG,A_3 \rangle \cQ +2 \langle \cG,A_3 \rangle  A_1+ \frac{4}{3} \langle \cG,A_2 \rangle \blp 2 \cG-A_2 \brp \non \\
&& \label{SectionEq3}  \\
0&=& I_4'(\cG,A_1,A_2)  +2 \langle \cG,A_1 \rangle \blp N \kappa \cG - A_2\brp + \langle \cG,A_2 \rangle \blp A_1-\cQ\brp \label{SectionEq4}\\
0&=& I_4'(\cG,A_1,A_1)  -2 \langle \cG,A_1 \rangle \cQ \,.   \label{SectionEq5}
\eea
We also need to the expansion of the single real constraint \eq{constraint2}:
\bea
\cO(r^4):\quad 0&=&2 \langle A_2,A_3\rangle -N\kappa  \langle \cG,A_3 \rangle  \label{ConstraintrOrder4}\\
\cO(r^3):\quad  0&=&  2 \langle A_1,A_3\rangle  +  \langle \cQ,A_3\rangle\label{ConstraintrOrder3}\\
\cO(r^2):\quad 0&=&  \langle A_1,A_2\rangle +N\kappa  \langle \cG,A_1 \rangle +  \langle \cQ,A_2\rangle \label{ConstraintrOrder2} \\
\cO(r^1):\quad 0&=&  2  \langle \cQ,A_1\rangle \label{ConstraintrOrder1} \,.
\eea
Note that once again, the highest order in $r$ components of \eq{cVEq4} is independent of the NUT charge and therefore the solution for $A_3$ can be taken from \cite{Halmagyi:2014qza}:
\be
A_3=\frac{1}{4}\frac{I_4'(\cG)}{\sqrt{I_4(\cG)}}\,,\qquad v_4 =\frac{1}{R^2_{{\rm AdS}_4}}=\sqrt{I_4(\cG)}\,.
\ee

We solve these equations with the ansatz
\bea
A_1&=&   a_{1} I_4'(\cG,\cG,\cG)+a_{2} I_4'(\cG,\cQ,\cQ)+a_{3} I_4'(\cG,\cQ,\cQ)+a_{4} I_4'(\cQ,\cQ,\cQ)  \\
A_2&=&   b_{1} I_4'(\cG,\cG,\cG)+b_{2} I_4'(\cG,\cQ,\cQ)+b_{3} I_4'(\cG,\cQ,\cQ)+b_{4} I_4'(\cQ,\cQ,\cQ) 
\eea
where $\{a_{i},b_j\}$ are real constants with non-trivial dependence on $(\cG,\cQ)$. The solution proceeds in much the same manner as in the work \cite{Halmagyi:2014qza} where the solution was found for $N=0$. The IR conditions which give $a_i$ in terms of $(\cG,\cQ)$ are the same we obtained for the case when $e^{2V}$ had a pair of double roots and are thus given by \eq{PairHorizon1}-\eq{PairHorizon4}.

Then from \eq{SectionEq4} we find the solution for $\{b_1,b_2,b_4\}$ in terms of $b_3$:
\bea
b_1&=&\frac{b_3 I_4(\cQ) \IGGGQ}{3I_4(\cG) \IGQQQ} - \frac{2 b_3 \IGQQQ}{3\IGGGQ} \non \\
&&+ \frac{b_3\kappa \IGGGQ \IGQQQ^2}{54 I_4(\cG)  \Pi_3} + \frac{N\kappa \IGQQQ^2}{18 \Pi_3} \\
b_2&=& \frac{\IGGGQ \blp 6N I_4(\cG) I_4(\cQ) - b_3 \Pi_2\brp }{6 I_4(\cG) \Pi_3} \\
b_4&=&-\frac{ \IGQQQ \blp 3 N I_4(\cG) + b_3 \IGGGQ \kappa \brp}{9\Pi_3}
\eea
Finally from \eq{SectionEq3} we solve for $b_3$ and find the rather lengthy expression
\be
b_3= \frac{b_n}{b_d}\non
\ee
where the numerator and denominator are given by
\bea
b_n&=& 6N\kappa I_4(\cG) \IGGGQ \IGQQQ^2 \IGGGQQQ \Pi_7  \non \\
&&+3\Bslb -I_4(\cG)^{3/2} \IGGGQ \IGQQQ \Pi_3^2 \Pi_8  \Bslb -18 I_4(\cG) \Pi_3^2\non \\
&& +(\kappa +4 N^2 I_4(\cG)^{1/2}) \IGGGQ^{1/2} \IGQQQ \Pi_5\non \\
&& -8N^2 I_4(\cG)^{3/2}  \bslb 144 \kappa I_4(\cQ)^2 \IGGGQ^2- \kappa \IGGGQ \IGQQQ^3\non \\
&&+72 I_4(\cQ) \IGQQQ \Pi_6 \bsrb \Bsrb  \Bsrb^{1/2}
\eea
and
\bea
b_d&=&  8I_4(\cG)\Bslb \IGGGQ\Bslb 2\kappa I_4(\cQ) \IGGGQ^2 (144 I_4(\cQ)^2\IGGGQ- \IGQQQ^3)  \non \\
&& \hspace{-1.5cm}+ \IGGGQ \IGQQQ (288 I_4(\cQ)^2 \IGGGQ -\IGQQQ^3)\IGGGQQQ   \non \\
&&\hspace{-1.5cm}+ 90\kappa I_4(\cQ) \IGGGQ \IGQQQ^2 \IGGGQQQ^2\non \\
&&\hspace{-1.5cm}+ 9 \IGQQQ^3  \IGGGQQQ^3 \Bsrb+ 18 \kappa I_4(\cG) \IGQQQ^2 \Pi_3 \Bsrb \non \\
&&\hspace{-1.5cm} -4\kappa \IGGGQ^3 \IGQQQ \Pi_5
\eea
We have used the notation
\bea
\Pi_1 &=& \IGQQQ \langle I'_4(\cG),I'_4(\cQ)\rangle  + 2\kappa \IGGGQ I_4(\cQ)\,,  \\
\Pi_2&=&  \IGGGQ \langle I'_4(\cG),I'_4(\cQ)\rangle+2\kappa \IGQQQ  I_4(\cQ)   \\
\Pi_3 &=& \IGQQQ \langle I'_4(\cG),I'_4(\cQ)\rangle  + 4\kappa \IGGGQ I_4(\cQ) \,, \\
\Pi_4&=& 2\kappa I_4(\cQ) \IGGGQ^2 + \IGQQQ \Pi_1  \\
\Pi_5 &=& \IGQQQ \langle I'_4(\cG),I'_4(\cQ)\rangle  + 2\kappa \IGGGQ I_4(\cQ) \,, \\
\Pi_6 &=& \IGGGQ \langle I'_4(\cG),I'_4(\cQ)\rangle  + 2\kappa \IGQQQ I_4(\cG) \,, \\
\Pi_7&=&2\kappa I_4(\cG) \IGQQQ^2 +\IGGGQ \Pi_5  \\
\Pi_8&=&2\kappa I_4(\cG) \IGGGQ^2 +\IGQQQ \Pi_6 \,.
\eea

These expression are fairly lengthy but in fact their derivation in Mathematica starting from \eq{SectionEq1}-\eq{ConstraintrOrder1} is quite straightforward when using the identities in appendix A.3 of \cite{Halmagyi:2014qza}. The $N\ra 0$ limit of these expressions agrees with those found in \cite{Halmagyi:2014qza}.

\section{Four Independent Roots}\label{sec:Independent}

While extremal black holes necessarily have a double real root in $e^{2V}$, more general configurations are possible. For example we could have one or two pairs of complex conjugate roots. A natural ansatz for such solutions is
 \bea
e^{2V} &=& v_0+v_1 r+ v_2r^2  + v_4 r^4  \label{FourInd1}\\
\Im \tcV&=& e^{-V} \hA   \label{FourInd2}\\
\hA&=&\Bslb  A_0+A_1 r + A_2 r^2 + A_3 r^3 \Bsrb\,.\label{FourInd3}
\eea
We have used a shift symmetry in $r$ to set $v_3=0$ but one cannot in general use a real shift in $r$ to set $v_0=0$.

An example of such solutions is the constant scalar asymptotically AdS$_4$ solution of section \ref{constantscalar}, corresponding to the STU-model with
\be
p^0=q_i=P\,,\qquad q_0=-p^i=Q\,.
\ee
In our formalism we find this constant scalar example to be given by the following data:
\bea
A_0&=&   \frac{N\kappa (P-1)}{2g} \cG + \frac{N\kappa }{48 g^3} I'_4(\cG,\cG,\cG) \\
A_1&=&   \frac{Q}{2g} \cG + \frac{P-3g N^2}{48 g^3} I'_4(\cG,\cG,\cG) \\
A_2&=&   \frac{N\kappa }{2} \cG \\
A_3&=& \frac{I'_4(\cG,\cG,\cG)}{24I_4(\cG)^{1/2}}
\eea
and the metric is given by
\bea
e^{2(V-U)}&=& r^2+N^2 \\
e^{2V}&=& 2\Bslb P^2+Q^2+g^2 N^4-2gN^2 P +4gN\kappa Qr +2(3gN^2-gP)r^2 +gr^4 \Bsrb  \,.
\eea
The phase of the spinor is given by
\bea
\sin \psi&=& e^{U-2V}\blp g r^3 +(-P+3g N^2 )r +N \kappa Q  \brp \,.
\eea

We have tried obtain generalizations of this solution using the ansatz \eq{FourInd1}-\eq{FourInd3} but have not managed to decouple the set of algebraic equations. However this should not be seen as evidence that such solutions do not exist. Such solutions would not necessarily correspond to black holes since that requires the existence of a horizon. Since we expect BPS black holes to have extremal horizons, these solutions are covered by our analysis in section \ref{sec:DoubleRoot}. Nonetheless looking ahead to possible extensions to Euclidean solutions, it is of some interest to have more general solutions with single real roots of $e^{2V}$.

\section{Conclusions}

We have solved quite generally the problem of AdS$_4$-NUT black holes in FI-gauged supergravity. There are various caveats we have spelt out in the text: our restriction to theories where $\cM_v$ is a symmetric very special K\"ahler manifold and that we have explicit solutions only when the charges obey the constraint \eq{AdS2Constraint} and thus the IR region is AdS$_2\times \Sig_g$. It may be possible to lift these restrictions but probably the more interesting direction would be to study BPS black holes with non-trivial rotation and acceleration parameters. Restricted classes of non-BPS solutions with rotation parameter have already been found \cite{Chong:2004na, Chow:2013gba} and it seems plausible that the methods used in this work can be adapted to include both rotation and acceleration. Very recently a solution with non-vanishing acceleration parameter and running scalars has been found \cite{Lu:2014ida}. 

It has been fleshed out in a nice series of works \cite{Martelli:2011fu, Martelli:2011fw, Martelli:2012sz, Martelli:2013aqa} that when continued to Euclidean signature, both the NUT and acceleration parameters correspond to the two squashing parameters of the $S^3$. It would be interesting to extend this interpretation to include more general bulk charges and thus more general ways of preserving supersymmetry on the boundary theory. Another direction which promises to be interesting is the study of half-BPS AdS$_4$-NUT black holes with running scalars. With more supersymmetry it could be fruitful to study individual microstates of such black holes. Central to the understanding of the boundary holographic interpretation of our solutions is to perform the Euclidean continuation and to compute the holographic free energy. We certainly hope to report on these interesting issues shortly.
 
The Janis-Newman algorithm is a solution-generating technique that was originally used for obtaining rotating solutions from a static seed metric using a complex coordinate transformation assorted with a specific modification of the metric functions \cite{Newman:1965tw,Newman:1965my}.
It was later extended by Demianski to add a NUT charge and to solutions with non-vanishing cosmological constant \cite{Demianski:1972uza}.
The latter transformation was obtained by solving equations of motion and Demianski did not give the rules for performing the transformation directly on the metric, implying that it was not possible to apply it in other cases.
Recently it was shown in \cite{Erbin:2014aja,Erbin:2015pla,Erbin:2014aya,Keane:2014sta} how to modify the prescription to accommodate the NUT charge and how to perform the transformation on gauge fields and on (complex) scalar fields, and for topological horizons.
As a consequence all the tools exist for application to gauged supergravity, and it would be very interesting to study how the black holes of the current paper can be obtained from the static black holes of~\cite{Halmagyi:2014qza}.
A first step in this direction has been taken by Klemm and Rabbiosi who proved that the NUT black hole from \cite{Gnecchi:2013mja} could be derived from the static black hole using this formalism \cite{privcommKlemm}.

\vskip 10mm
\noindent{\bf Acknowledgements} We would like to thank Guillaume Bossard, Davide Cassani and Dietmar Klemm for discussions. This work was conducted within the framework of the ILP LABEX (ANR-10-LABX-63) supported by French state funds managed by the ANR within the Investissements d'Avenir programme under reference ANR-11-IDEX-0004-02.

\vskip 10mm

\begin{appendix}

\section{Special Geometry Background}\label{app:Special}

Here we summarize our conventions from special geometry

\subsection{Generalities}

We will use the conventions where the prepotential is given by
\be
\cF=-\frac{d_{ijk}X^i X^j X^k}{X^0}\,,
\ee
the metric is 
\be
g_{ij}=-\frac{3}{2}\frac{d_{y,ij}}{d_y} + \frac{9}{4}\frac{d_{y,i}d_{y,j}}{d^2_y}\,.
\ee
where
\be
X^\Lam=\bpm1 \\ z^i \epm\,,\quad\quad z^i=x^i+iy^i
\ee
and the covariant tensor is given by
\be
\hd^{ijk} = \frac{g^{il} g^{jm} g^{kn}d_{ijk}}{d_y^2}\,.
\ee
When $\cM_v$ is a symmetric very special K\"ahler manifolds the tensor $\hd^{ijk}$ has constant entries and satisfies
\be
\hd^{ijk}d_{jl(m}d_{mp)k} =\frac{16}{27} \Bslb \delta^i_{l} d_{mnp} + 3 \delta^i_{(m} d_{mp)l} \Bsrb\,.
\ee
The sections are given by
\be
\cV=\bpm L^\Lam \\ M_\Lam \epm \ =\ e^{K/2} \bpm X^\Lam \\ F_\Lam  \epm \label{cVdef}
\ee
and satisfy\footnote{The symplectic inner product is $\langle A,B\rangle= B^\Lam A_\Lam-B_\Lam A^\Lam$.}
\be
\langle \cV,\cVbar \rangle = -i\,,\quad\quad \langle D_i\cV,D_{\jbar}\cVbar \rangle=ig_{i\jbar}
\ee
and any symplectic vector can be expanded in these sections. For example the charges are expanded as
\be
\cQ= i \cZbar \cV - i \cZ \cVbar + i \cZbar^{\ibar} D_{\ibar} \cVbar - i \cZbar^i D_i\cV
\ee
where 
\be
\cZ=\langle \cQ,\cV\rangle\,,\quad\quad \cZ_i = \langle \cQ,D_i \cV \rangle\,.
\ee
We also have a complex structure on the symplectic bundle over $\cM_v$:
\be
\Om\cM \cV= -i \cV\,,\quad\quad \Om\cM (D_i \cV)= i D_i \cV
\ee
where 
\be
\Om=\bpm0 & -1\!\!1 \\ 1\!\! 1 & 0 \epm\,,\quad\quad\cM=\bpm 1 & -\cR \\ 0 & 1 \epm\bpm \cI & 0 \\ 0 & \cI^{-1} \epm\bpm 1 & 0 \\ -\cR & 1 \epm
\ee
and $\cN=\cR+i\cI$ is the standard matrix which gives the kinetic and topological terms in the action for the gauge fields.

\subsection{The Quartic Invariant} \label{app:quartic}

Symmetric very special K\"ahler manifolds were classified in \cite{Gunaydin:1983bi, Cremmer:1984hc} (see also \cite{Cecotti:1988ad, deWit:1991nm, deWit:1992wf, deWit:1993rr, deWit:1995tf}) and there is one infinite family as well as several sporadic cases.
For each of these manifolds one can define the quartic invariant:
\bea
I_4(\cQ)&=& \frac{1}{4!}  t^{MNPQ}\cQ_M \cQ_N \cQ_P \cQ_Q \non \\
&=&-(p^0q_0+p^iq_i)^2 -4q_0 d_{ijk} p^ip^jp^k +\frac{1}{16}p^0 \hd^{ijk}q_iq_jq_k + \frac{9}{16} d_{ijk} \hd^{ilm} p^j p^kq_l q_m\,.
\eea
Recall that the indices take values $\Lam=0\,,\ldots\,,n_v$ and $i=1\,,\ldots\,, n_v$. Then the indices $\{M,N,P,Q\}$ take  both $\Lam$ indices up and own, for example
\be
\cQ _M= \bpm p^\Lam \\ q_\Lam \epm\,.
\ee
Symmetric spaces are cosets $G/H$ and $I_4(\cQ)$ is invariant under the global symmetries $G$ of the coset. In the work of de Wit and Van Proeyen one can find the explicit embedding of $G$ into the symplectic group $Sp(2n_v+2,\RR)$ which then acts in a straightforward manner on $I_4(\cQ)$. The first steps incorporating this quartic invariant into the lexicon of BPS black holes were taken in \cite{Cvetic:1995bj, Kallosh:1996uy, Cvetic:1996zq} it is quite remarkable how integral it has become. Some more recent references which utilize it are \cite{Cerchiai:2009pi, Ferrara:2011di, Bossard:2012xsa}.

Using the four index tensor $t^{MNPQ}$ one can also define $I_4$ evaluated on four distinct symplectic vectors as well as its derivative $I'_4$ which is itself a symplectic vector. We essentially use the same normalizations as in\footnote{Note that $\Omega_{MN}$ are the components of $\Omega^{-1} = - \Omega$.} \cite{Katmadas:2014faa}:
\bea
I_4(A,B,C,D)&=& t^{MNPQ}A_M B_N C_P D_Q \\
I'_4(A,B,C)_M&=& \Om_{MN}t^{NPQR}A_P B_Q C_R \\
I_4(A)&=& \frac{1}{4!}  t^{MNPQ}A_M A_N A_P A_Q \\
I'_4(A)_M&=& \frac{1}{3!}  \Om_{MN}t^{NPQR}  A_P A_Q A_R
\eea
We then have
\be
24 I_4(A)= I_4(A,A,A,A)\,,\quad\quad 6 I'_4(A)= I'_4(A,A,A)\,.
\ee
A useful identity which is valid when $\cM_v$ is a symmetric space and which plays the key role in deriving the form of the BPS equations given in section \ref{sec:Symmetric} is
\be\label{I4pQImVImV}
I'_4(A,\Im\cV,\Im\cV)= 4 \Im\bslb \langle A,\cV \rangle\bsrb \Im \cV + 8 \Re\bslb \langle A,\cV \rangle\bsrb \Re \cV  - \Om \cM A\,.
\ee
Using this identity and replacing $A \ra \Im \cV$ we derive the useful expressions
\be\label{ReVImV}
\Re \cV = -\frac{I'_4(\Im\cV)}{2\sqrt{I_4(\Im\cV)}}\,,\qquad
I_4(\Im\cV) = \frac{1}{16}\,.
\ee

\subsection{The STU Model}\label{app:STU}

Our gauged supergravity theories are specified by a choice of scalar manifold $\cM_v$, $\cV$, and a symplectic vector of gauging parameters $\cG$. The so-called STU-model is the model where the sections are derived from the prepotential
\be
F=-\frac{X^1X^2 X^3}{X^0} \label{STUprepot}
\ee
or one of its symplectically dual frames. Taking the prepotential \eq{STUprepot} and the vector of gauging parameters 
\be
\cG=\bpm g^\Lam \\ g_\Lam \epm \,,\qquad g^\Lam = -  \bpm 0\\g \\g \\g \epm\,,\qquad g_\Lam =\bpm  g \\ 0\\0 \\0 \epm \label{STUGaugings}
\ee
we can embed this into the de Wit-Nicolai $\cN=8$ theory \cite{deWit:1982ig} using the results of \cite{Cvetic:1999xp}. Throughout this article we will refer this particular STU model as the ``STU-model". Using a simple symplectic transformation, this is equivalent to a frame where 
\be
F=-2i\sqrt{X^0X^1X^2X^3}\,,\qquad g^\Lam = 0 \,,\qquad g_\Lam =\bpm g\\g \\g \\g \epm
\ee
which is the frame found in \cite{Cvetic:1999xp}.

\section{The Derivation of the BPS Equations} \label{app:BPSDerivation}

\subsection{Metric}

We follow as much as possible the conventions of \cite{Andrianopoli:1996vr},in particular the signature is $(+---)$. The metric is
\be
e^0=e^{U} (dt+2Nf' (\tha) d\phi)\, ,\ \  e^1 =e^{-U} dr\, , \ \
 e^2= e^{V-U} d\tha \,,\ \  e^3= e^{V-U} f( \tha) d\vphi 
\ee
with the different horizons labelled by
\be
f(\tha)=\left\{ \barr{ll} \sin \tha& S^2\quad (\kappa=1) \\ \sinh \tha & \HH^2\quad (\kappa=-1) \earr \right. 
\ee
We will derive here the equations for $\kappa=\pm$ and discuss the case of $\kappa=0$ at the end of this section. We note that $f''(\tha)= -\kappa f(\tha)$ and the spin connection is
\bea
\om_{001}&=&U'e^U\,, \qquad \om_{023}=  N\kappa e^{3U-2V} \non \\
 \om_{212}&=&(V'-U')e^U\,, \qquad \om_{203} = N\kappa e^{3U-2V}\non\\
\om_{313}&=&(V'-U')e^U\,,\ \ \om_{323}=e^{U-V}\frac{f'}{f}\qquad \om_{302} = -N \kappa e^{3U-2V}\ \,.\non
\eea
%
\subsection{Gauge Field}

The gauge field strengths are
\be
F^\Lam_{\mu\nu}=\half (\del_\mu A^\Lam_{\nu}-\del_{\nu}A^\Lam_{\mu})
\ee
We take 
\bea
F^\Lam&=&  \tq'^\Lam dr\w (dt+2N f' d\phi )+   P^\Lam f d\tha \w d\phi  \,,
\eea
the Bianchi identity sets
\bea
&&  P'^\Lam+ 2\kappa N  \tq'^\Lam=  0 \label{pPrimezero}\,.
\eea
A representative of the gauge potential is
\bea
A^\Lam &=& 
  \tq^\Lam dt -\kappa P^\Lam f' d\phi 
\eea

\subsubsection{Maxwell's Equation}

The dual gauge field strength is
\be
G_{\Lam}= \cR_{\Lam\Sig} F^\Sig - \cI_{\Lam\Sig} *F^{\Sig}
\ee
So that
\bea
0=dG_\Lam
\eea
gives
\bea
0
&=& \Bslb  P^\Lam \cR_{\Lam\Sig}     -  \cI_{\Lam\Sig}\tq'^\Lam e^{2(V-U)} \Bsrb' + 2N \kappa e^{2(U-V)} \bslb \blp\cI  + \cR \cI^{-1} \cR\brp_{\Lam\Sig}  P^\Lam  -(\cI^{-1}\cR)^{\Lam}_{\ \Sig} Q_\Lam \bsrb 
\eea 
So we define 
\be
Q_\Lam =- \cI_{\Lam\Sig}\tq'^\Sig e^{2(V-U)} +P^\Sig \cR_{\Lam\Sig} 
\ee
then in combination with \eq{pPrimezero} we get the symplectic invariant Maxwell's equations
\be
\hcQ'= -2N\kappa e^{2(U-V)}\Om \cM \hcQ
\ee
where
\be
\hcQ= \bpm  P^\Lam \\ Q_\Lam \epm\,. \label{MaxEq1}
\ee
The Maxwell fields are sourced by the NUT charge. 
The contraction of the charges with the section is denoted
\bea
\hcZ=\langle \hcQ, \cV \rangle \,,\qquad\hcZ_i=\langle \hcQ, D_i \cV \rangle 
\eea

\subsection{Gravitino Variation}

The gravitino variation is
\bea
\delta \psi_\mu &=& \cD_\mu \eps_A + i S_{AB} \gam_\mu \eps^B + 2i \cI_{\Lam \Sig} \cF^{-\Lam}_{\mu\nu} \gam^\nu \eps_{AB} \eps^B 
\eea
where the various quantities are defined as follows
\bea
\cD_\mu \eps_A &=& \hD_\mu\eps_A  + \frac{i}{2}(\sig^3)_A^{\ B}A_\mu^\Lam g_\Lam \eps_B \\
 \hD_\mu\eps_A &=& D_\mu \eps_A + \frac{i}{2}A_\mu \eps_A \\
S_{AB}&=& \frac{i}{2} (\sig^3)_A^{\ C} \eps_{BC} g_\Lam L^\Lam \\
A_\mu&=& \frac{1}{2i} (K_i \del_\mu z^i - K_{\ibar} \del_\mu \zbar^{\ibar})
\eea 

The gravitino variations for our ansatz are
\bea
 0&=&\gam^0\del_0\eps_A
+\frac{U'e^U}{2}\gam^{1}\eps_A -\frac{i}{2}  N\kappa e^{3U-2V}  \gam^1\eps_A
+\frac{i}{2}  e^{-U}\tq^\Lam  g_\Lambda \, \gam^0(\sigma^3)_A^{\ B} \epsilon_B +iS_{AB}\eps^B \non \\
&& +\frac{i}{2} e^{2(U-V)} \gam^{01} \hcZ  \eps_{AB} \eps^B \label{gr1} \\
0&=&\gam^1\hD_1\eps_A +i S_{AB}\eps^B + \frac{i}{2} e^{2(U-V)} \gam^{01} \hcZ  \eps_{AB} \eps^B \label{gr2} \\
0&=&\gam^2\del_2\eps_A
+ \half (V'-U')e^U \gam^1\eps_A+\frac{i}{2}  N\kappa e^{3U-2V} \gam^1\eps_A
+i S_{AB}\eps^B    -\frac{i}{2}  e^{2(U-V)} \hcZ\gam^{01}  \eps_{AB} \eps^B \label{gr3}\\
0&=&\gam^3 \del_3\eps_A
+\half e^{U-V} \frac{f'}{f} \gam^2 \eps_A+\half (V'-U')e^U \gam^1 \eps_A + \frac{i}{2}   N\kappa e^{3U-2V} \gam^1\eps_A \non \\
&& -\frac{i}{2}  e^{U-V} \frac{\kappa f'}{f} (P^\Lam+2\tq^\Lam N \kappa) g_\Lambda \, \gam^3(\sig^3)_A^{\ B} \epsilon_B  +iS_{AB}\eps^B -\frac{i}{2}  e^{2(U-V)}\hcZ\gam^{01}   \eps_{AB} \eps^B 
  \label{gr4}
\eea

\subsection{Radial Dependence}

Taking \eq{gr1}-\eq{gr2} we get
\bea
D_1 \eps_A 
&=&\frac{e^U}{2}\blp U' -iN\kappa e^{2(U-V)}\brp \eps_A
+\frac{i}{2}  e^{-U}\tq^\Lam  g_\Lam \, \gam^{01}(\sig^3)_A^{\ B} \epsilon_B   \non \\
\Rightarrow\quad \del_r \eps_A &=& -\frac{i}{2}A_r \eps_A +\frac{1}{2}\blp U' -iN\kappa e^{2(U-V)}\brp \eps_A
+\frac{i}{2}  e^{-2U}\tq^\Lam  g_\Lam \, \gam^{01}(\sig^3)_A^{\ B} \epsilon_B \non
\eea
With the spinor ansatz 
\be
\eps_A= e^{(H+i\psi)/2} \eps_{0A}
\ee
we get 
\bea
H'&=&U'  \\
(\psi' +A_r+N\kappa e^{2(U-V)}) \eps_A &=&  e^{-2U}\tq^\Lam  g_\Lam \, \gam^{01}(\sig^3)_A^{\ B} \epsilon_B \label{psiDepend}
\eea

\subsection{Deriving the Projectors}

Taking \eq{gr1}-\eq{gr3} gives one of the projectors
\bea
0&=&i \tq^\Lam g_\Lam (\sig^3)^{\ B}_A e^{-U}\gam^0\eps_B 
+(2U'-V'-2i  N\kappa e^{2(U-V)} )e^U\gam^{1}\eps_A  + 2 ie^{2(U-V)}  \gam^{01} \hcZ \,  \eps_{AB} \eps^B\label{BPS4}
\eea
and the remaining $\tha$-dependent pieces in \eq{gr4} should be separately set to zero which is the usual statement of {\it setting the gauge connection equal to the spin connection}, giving the other projector
\bea
 \eps_A&=& -    \kappa (P^\Lam +2N\kappa\tq^\Lam) g_\Lam \,(\sig^3)_A^{\ B} \gam^{01} \epsilon_B \label{firstproj}
\eea
The integrability condition for this projector is:
\be
1=  \blp(P^\Lam+2N\kappa \tq^\Lam)  g_\Lam\brp^2 
\ee
which we solve with
\be\fbox{$
(P^\Lam+2N\kappa \tq^\Lam)g_\Lam = \veps_p \kappa
$}\label{DiracBPS}
\ee
As a result we have the first projector \eq{firstproj} to be
\be\fbox{$
\eps_A=    - \veps_p  (\sigma^3)_A^{\ B} \gam^{01} \epsilon_B 
$}\ee
with $\veps_p=\pm1$ is a sign choice which we will retain.

The remaining equation is from \eq{gr4}
\bea
 S_{AB}\eps^B&=&\half \gam^{01} e^{2(U-V)}\hcZ  \eps_{AB} \eps^B+\frac{i}{2}  (V'-U' + i  N\kappa e^{2(U-V)} )e^U \gam^1  \eps_A \label{BPS5}
\eea
which gives
\bea
\eps_A &=&\frac{ 2ie^{U-2V}\hcZ }{ (2U'-V'-2i  N\kappa e^{2(U-V)} -i\veps_p e^{-2U} \tq^\Lam g_\Lam )  } \eps_{AB}\gam^{0} \eps^B  \label{epsproj2}\,.
\eea
We want the consistency condition arising from this, which is
\be
2e^{U-2V}e^{-i\tha} \hcZ   =  \sqrt{|2U'-V'-2i  N\kappa e^{2(U-V)}-i\veps_p e^{-2U} \tq^\Lam g_\Lam|^2 }  \label{Nphase}\ee
 where we take the positive sign in the square root. Then the projector is
 \be 
 \eps_A=\frac{ie^{i\tha} \sqrt{ |2U'-V'-2i  N\kappa e^{2(U-V)}-i\veps_p e^{-2U} \tq^\Lam g_\Lam|^2}}{ (2U'-V'-2i  N\kappa e^{2(U-V)} -i\veps_p e^{-2U} \tq^\Lam g_\Lam )} \eps_{AB}\gam^{0}  \eps^B 
 \label{proj3}\ee
 We define the phase
 \be
e^{i\psi}=e^{-i\tha}\frac{2U'-V'- 2i  N\kappa e^{2(U-V)} - i\veps_p e^{-2U} \tq^\Lam g_\Lam }{ \sqrt{ |2U'-V'-2i  N\kappa e^{2(U-V)}-i\veps_p e^{-2U} \tq^\Lam g_\Lam |^2}} \label{psi}
 \ee
so that the projector \eq{proj3} is
 \be\fbox{$ \barr{rcl}
 \eps_A&=&ie^{i\psi}  \eps_{AB}\gam^{0}  \eps^B  \\
 \eps^A&=&ie^{-i\psi}  \eps^{AB}\gam^{0}  \eps_B  
\earr $}\label{proj4}\ee
and \eq{Nphase} becomes
\be\fbox{$
2e^{U-2V} e^{-i\psi} \hcZ = 2U'-V' -2i  N\kappa e^{2(U-V)}- i\veps_p e^{-2U} \tq^\Lam g_\Lam 
$}\label{BigEq1}
\ee
 
We now examine \eq{BPS5}, it gives the projector
\bea
 S_{AB}\eps^B
&=&  \frac{\veps_p}{2}\Bslb- e^{2(U-V)}\hcZ   - (V'-U'+i  N\kappa e^{2U-2V})e^{U}e^{i\psi} \Bsrb  (\sig^3)_{AB}\eps^B
\eea
which gives the bosonic constraint
\be\fbox{$
i \veps_p e^{-i\psi}\cL=- e^{2(U-V)}e^{-i\psi} \hcZ  +e^{U}(U'-V'-iN\kappa e^{2U-2V})
$} \label{BigEq2}\ee

In summary, the final equations coming from \eq{BigEq1} and \eq{BigEq2} are
\bea
 e^U U'&=&\veps_p \Im( e^{-i\psi} \cL)+  e^{2(U-V)}\Re (\hcZ e^{-i\psi})   \label{gravEqs1}   \\
 e^U V'  &=&  2 \veps_p \Im( e^{-i\psi}\cL) \\
 0&=&N\kappa e^{3U-2V} + \veps_p\Re(e^{-i\psi} \cL) + e^{2(U-V)}\Im ( e^{-i\psi}\hcZ)   \label{gravEqs3} \\
 0&=& 2\Im(e^{-i\psi} \hcZ)+2 N\kappa e^U+\veps_p e^{2V-3U} \tq^\Lam g_\Lam  \label{gravEqs4}
\eea
In addition we get the radial dependence from \eq{psiDepend} to be
\bea
\psi' &=&  -A_r   -N\kappa e^{2(U-V)}  -\veps_p   e^{-2U}\tq^\Lam  g_\Lam   \label{gravEqs5} 
\eea
and the Dirac quantization condition \eq{DiracBPS}
\be
(P^\Lam+2N \tq^\Lam)g_\Lam = \veps_p \kappa\,. \label{gravEqs6} 
\ee
\subsection{Gaugino Variation}

The gaugino variation is
\bea
0&=& i \nabla_\mu z^i \gam^\mu \eps^A + \Blp  -g^{i\jbar} \fbar^{\Sigma}_{\jbar} \cI_{\Sig \Lam} \cF^{- \Lambda}_{\mu\nu} \gamma^{\mu\nu }\eps^{AB}  + W^{i A B} \Brp\eps_B 
\eea
where
\be
W^{iAB}= i(\sig^3)_C^{\ B}\eps^{CA} g_\Lam g^{i\jbar} \fbar^\Lam_{\jbar}
\ee
We use \eq{proj4} to get
\bea
0&=&   - \veps_p ( e^{-i\psi} e^{U}\dot{z}^i-e^{2(U-V)}\hcZ^i) -ig^{i\jbar} {\bar f}_{\jbar}^{\Lam}g_\Lam 
\eea
and arrive at
\be\fbox{$
 e^{-i\psi} e^{U}\dot{z}^i   = e^{2(U-V)} \hcZ^i  -\veps_p i   g^{i\jbar} {\bar f}_{\jbar}^{\Lambda} P_\Lam^3
$}\label{gauge3}
\ee

We can simplify \eq{gauge3} a bit further by contracting with $f_i^\Lam$. In principle this could lose equations since $f_i^\Lam$ cannot be invertible: 
\bea
e^{-i\psi} e^{U}  \dot{z}^i f_i^\Delta& =&- f_i^\Delta g^{i\jbar} \fbar^{\Lam}_{\jbar}\Bslb - e^{2(U-V)}\blp Q_\Lam - (\cR+i \cI)_{\Lam \Sig} P^\Sig \brp+  i \veps_p    g_\Lam \Bsrb
\eea
and after some algebra we get
\be
\fbox{$\barr{rcl}
\del_r  \blp e^{U} \Re \cL^\Delta \brp   & =& \frac{1}{2} e^{2(U-V)}\Blp\cI^{\Delta\Lam}\cR_{\Lam \Sig} P^\Sig - \cI^{\Delta \Lam} Q_\Lam\Brp \\
\del_r \blp e^{-U}\Im \cL ^\Delta \brp& =&\frac{ P^{\Delta}}{2e^{2V}} + \frac{ \veps_p }{2 e^{2U }}   \cI^{\Delta \Sig}  g_\Sig  +2(\veps_p   e^{-2U}\tq^\Lam  g_\Lam+ N \kappa  e^{U-2V}) \Re \cL^\Delta
\earr $}
\label{sectionEqs1}\ee
where we have defined the section with a phase
\be
\cL^\Lam = e^{-i\psi} L^\Lam\,.
\ee
It is also useful to have a derivation of the equation for the other component of the sections
\be
M_\Lam = \cN_{\Lam\Sig} L^\Sig
\ee
and we get
\bea
e^{-i\psi} e^{U}  \dot{z}^i h_{i\Upsilon}& =&- \cNbar_{\Upsilon \Delta}f_i^\Delta g^{i\jbar} \fbar^{\Lam}_{\jbar}\Bslb - e^{2(U-V)}\blp Q_\Lam - (\cR+i \cI)_{\Lam \Sig} P^\Sig \brp+  i \veps_p    g_\Lam \Bsrb \non \\
LHS&=&e^{U}   \del_r \cM_\Upsilon  -i e^{U} (\veps_p   e^{-2U}\tq^\Lam  g_\Lam+N \kappa e^{2(U-V)})  \cM_\Upsilon  \\
RHS& =& (\cR-i \cI)_{\Upsilon \Delta}\Bslb- \half e^{2(U-V)}   \cI^{\Delta \Lam} \blp Q_\Lam - (\cR+ i \cI)_{\Lam \Sig} P^\Sig \brp
 +\frac{ i\veps_p }{2}   \cI^{\Delta \Sig}  g_\Sig  \non \\
&& + \cLbar^{\Delta} \bslb -e^U U'  +iN\kappa e^U+ i\veps_p e^{2V-3U} \tq^\Lam g_\Lam  \bsrb\Bsrb \non
\eea
This gives
\be
\fbox{$\barr{rcl}
\del_r  \blp e^{U} \Re \cM_\Lam \brp   & =&  \frac{1}{2}e^{2(U-V)} \Bslb (\cI + \cR^{-1} \cI^{-1} \cR)_{\Lam \Sig} P^\Sig - (\cR \cI^{-1})_{\Lam}^{\ \Sig}Q_\Sig\Bsrb
  \\
\del_r \blp e^{-U} \Im \cM_{\Lam} \brp& =& \Bslb  8 \veps_p e^{2(V-U)} \Re\blp e^{-i\psi} \cL\brp + 4N \kappa e^{U} \Bsrb\Re \blp e^{-i\psi}\cM_\Lam \brp + Q_\Lam +\veps_p e^{2(V-U)}  (\cR \cI^{-1})_{\Lam}^{\ \Sig}g_\Sig  \non \\
\earr $}
\label{sectionEqs2}
\ee
where again we have defined the sections rescaled by a phase:
\be\cM_\Lam = e^{-i\psi} M_\Lam\,.
\ee

\subsection{Symplectic Covariant Equations}

While we have worked in a formalism with only electric gaugings $g_\Lam$ we can provide a symplectic completion of the BPS variations as follows by introducing a symplectic vector of gaugings 
\be
\cG=\bpm g^\Lam \\ g_\Lam \epm
\ee
We find that the following equations (recall that $\cV$ is defined in \eq{cVdef}), when rotated to a symplectic frame with only electric gaugings, are equivalent to the set comprised of \eq{MaxEq1},\eq{DiracBPS},\eq{gravEqs1}-\eq{gravEqs6}, \eq{sectionEqs1} and \eq{sectionEqs2}:
\bea
2 e^{2V} \del_r \Bslb \Im\blp e^{-i\psi}e^{-U}\cV \brp \Bsrb &=& \Bslb  8 \veps_p e^{2(V-U)} \Re\blp e^{-i\psi} \cL\brp + 4N \kappa e^{U} \Bsrb\Re \blp e^{-i\psi}\cV \brp + \hcQ -\veps_p e^{2(V-U)} \Om \cM \cG  \non \\
&& \label{FinalEq1}\\
2\del_r \bslb\Re ( e^Ue^{-i\psi}\cV)\bsrb&=& e^{2(U-V)} \Om \cM \hcQ +\veps_p \cG \label{FinalEq2}\\
\del_r \blp e^V \brp &=& 2 \veps_p e^{V-U} \Im \blp  e^{-i\psi} \cL\brp \label{FinalEq3} \\
\psi'+A_r&=&-2 \veps_p e^{-U}\Re (e^{-i\psi} \cL) - N\kappa e^{2(U-V)} \label{FinalEq4} \\
\hcQ'&=& -2N\kappa e^{2(U-V)} \Om \cM \hcQ \label{FinalEq5} \\
0&=&N\kappa e^{3U-2V} + \veps_p\Re(e^{-i\psi} \cL) + e^{2(U-V)}\Im ( e^{-i\psi}\hcZ)  \label{FinalEq6} \\
\veps_p \kappa&=& \langle \cG , \hcQ+4N \kappa  e^U \Re (e^{-i\psi}\cV )\rangle\,.\label{FinalEq7}
\eea
In fact we had two constraints \eq{gravEqs3} and \eq{gravEqs4} but one linear combination of them is implied by \eq{FinalEq2} and \eq{FinalEq5} up to a symplectic vector of constants. This is fleshed out in section \ref{sec:BPSEqs} where we will examine \eq{FinalEq1}-\eq{FinalEq7} in more detail.

Since we have worked from the outset with $\kappa=\pm 1$ we now explain how to obtain the equations for $\kappa=0$. This requires some rather straightforward modifications of our analysis, the outcome is that for $\kappa=0$ one should send $N\kappa\ra N$ in \eq{FinalEq1}, \eq{FinalEq4}, \eq{FinalEq5}, \eq{FinalEq6} and \eq{FinalEq7}, then set $\kappa=0$ in \eq{FinalEq7}.

\section{Identities for the Quartic Invariant}\label{app:identities}

In the derivation of our solution, we have used various identities for the quartic invariant which we tabulate here.

The formulas given in this appendix are a consequence of the Jordan algebra's structure of very special geometry, and the fact that the duality groups are of $E_7$-type~\cite{Brown:1969} (see also~\cite{Ferrara:2012qp, Bossard:2013oga} and references therein).
While some of them are proved in the above references, they have been determined by matching both sides on Mathematica.

\subsection{Order 5}

\begin{align*}
	I'_4(I'_4(A),A,A) &= 
		- 8\, A I_4(A) \\
	I'_4(I'_4(A),A,B) &= 2\, I'_4(A) \langle A, B \rangle 
		- \frac{1}{3}\,  A I_4(A,A,A,B) \\
	I'_4(I'_4(A,A,B),A,A) &= 
		- \frac{4}{3}\,  A I_4(A,A,A,B) 
		- 8\, I'_4(A) \langle A, B \rangle 
		- 16\, B I_4(A) \\
	I'_4(I'_4(A,A,B),A,B) &= 
		- \frac{1}{3}\,  2\, B I_4(A,A,A,B) 
		- 2\, A I_4(A,A,B,B) \non \\
		&\hspace{3cm}
		+ 2\, I'_4(A,A,B) \langle A, B \rangle 
		- 2\, I'_4(I'_4(A),B,B) \\
	I'_4(I'_4(A,B,B),A,A) &= 
		- \frac{4}{3}\,  B I_4(A,A,A,B) 
		- 4\, I'_4(A,A,B) \langle A, B \rangle 
		+ 2\, I'_4(I'_4(A),B,B)
\end{align*}

\subsection{Order 6}

\begin{align*}
	\langle I'_4(A,A,B), I'_4(A)\rangle &= 
		- 8\, I_4(A) \langle A, B \rangle \\
	\langle I'_4(A,B, B),I'_4(A) \rangle &= 
		- \frac{2}{3}\,  I_4(A,A,A,B) \langle A, B \rangle \\
	\langle I'_4(A,B,B), I'_4(A,A,B) \rangle &= 12\, \langle I'_4(A),I'_4(B) \rangle
		- 4\, I_4(A,A,B,B) \langle A, B \rangle
\end{align*}

\subsection{Order 7}

\begin{align*}
	I'_4(I'_4(A),I'_4(A),A) &= 8\, I_4(A) I'_4(A) \\
	I'_4(I'_4(A),I'_4(A),B) &= 4\, I_4(A) I'_4(A,A,B) 
		- \frac{2}{3}\,  I'_4(A) I_4(A,A,A,B)
		- 16\, A I_4(A) \langle A, B \rangle \\
	I'_4(I'_4(A),I'_4(A,A,B),A) &= 2\, I'_4(A) I_4(A,A,A,B)
		+ 16\, A I_4(A) \langle A, B \rangle \\
	I'_4(I'_4(A),I'_4(A,B,B),A) &= 2\, I'_4(A) I_4(A,A,B,B) 
		+ \frac{4}{3}\,  A I_4(A,A,A,B) \langle A, B \rangle \\
	I'_4(I'_4(A),I'_4(A,A,B),B) &= 8\, I_4(A) I'_4(A,B,B)
		- 2\, I'_4(A) I_4(A,A,B,B)  \\
		&+ \frac{1}{3}\,  I_4(A,A,A,B) I'_4(A,A,B) 
		- 16\, B I_4(A) \langle A, B \rangle  \\
		&- \frac{8}{3}\,  A I_4(A,A,A,B) \langle A, B \rangle \\
	I'_4(I'_4(A,A,B),I'_4(A,A,B),A) &= 
		- 16\, I_4(A) I'_4(A,B,B)
		+ 8\, I'_4(A) I_4(A,A,B,B)  \\
		&+ \frac{4}{3}\,  I_4(A,A,A,B) I'_4(A,A,B) 
		+ 64\, B I_4(A) \langle A, B \rangle \\
		&+ \frac{16}{3}\,  A I_4(A,A,A,B) \langle A, B \rangle \\
	I'_4(I'_4(A),I'_4(B),A) &= \frac{1}{3}\,  I'_4(A) I_4(A,B,B,B)
		+ 2\, A \langle I'_4(A),I'_4(B)\rangle \\
	I'_4(I'_4(A),I'_4(A,B,B),B) &= 
		- \frac{2}{3}\,  I'_4(A) I_4(A,B,B,B) 
		+ \frac{1}{3}\,  I_4(A,A,A,B) I'_4(A,B,B) \\
		&\qquad- \frac{4}{3}\,  B I_4(A,A,A,B) \langle A, B \rangle 
		- 8\, A \langle I'_4(A),I'_4(B)\rangle 
		+ 16\, I_4(A) I'_4(B) \\
	I'_4(I'_4(A,A,B),I'_4(A,A,B),B) &= 
		- \frac{16}{3}\,  I'_4(A) I_4(A,B,B,B) 
		+ \frac{8}{3}\,  I_4(A,A,A,B) I'_4(A,B,B) \\
		&\qquad- 16\, A I_4(A,A,B,B) \langle A, B \rangle 
		- \frac{16}{3}\,  B I_4(A,A,A,B) \langle A, B \rangle  \\
		&\qquad+ 32\, A \langle I'_4(A),I'_4(B)\rangle 
		+ 32\, I_4(A) I'_4(B) \\
	I'_4(I'_4(A,A,B),I'_4(A,B,B),A) &= \frac{16}{3}\,  I'_4(A) I_4(A,B,B,B)
		+ 2\, I_4(A,A,B,B) I'_4(A,A,B)  \\
		&\qquad- \frac{2}{3}\,  I_4(A,A,A,B) I'_4(A,B,B) 
		+ \frac{16}{3}\,  B I_4(A,A,A,B) \langle A, B \rangle  \\
		&\qquad+ 8\, A I_4(A,A,B,B) \langle A, B \rangle 
		- 8\, A \langle I'_4(A),I'_4(B)\rangle 
		- 32\, I_4(A) I'_4(B)
\end{align*}

\end{appendix}
\providecommand{\href}[2]{#2}\begingroup\raggedright\endgroup

\end{document}